\documentclass[journal]{IEEEtran}
\usepackage{graphicx}
\usepackage{multirow}
\usepackage{url}
\usepackage{listings,amsmath} 
\usepackage{ragged2e}
\usepackage{color}
\usepackage{marvosym}
\usepackage{cite}
\usepackage{subfigure}
\usepackage[colorlinks, linkcolor=black]{hyperref}

\ifCLASSINFOpdf
\else
\fi

\begin{document}
%
\title{Block Convolution: Towards Memory-Efficient Inference of Large-Scale CNNs on FPGA}
%
%
%

\author{Gang Li\textsuperscript{*}, Zejian Liu\textsuperscript{*}, Fanrong Li\textsuperscript{*},~\IEEEmembership{Student~Member,~IEEE}
        and~Jian Cheng\textsuperscript{\Letter},~\IEEEmembership{Member,~IEEE}
\thanks{*Equal contributions.}
\thanks{\Letter Corresponding author.}
\thanks{G. Li is with the Institute of Automation, Chinese Academy of Sciences, Beijing 100190, China, and the School of Artificial Intelligence, University of Chinese Academy of Sciences, Beijing 100049, China (gangli0426@gmail.com).}
\thanks{Z. Liu and F. Li are with the Institute of Automation, Chinese Academy of Sciences,
Beijing 100190, China, and the School of Future Technology, University
of Chinese Academy of Sciences, Beijing 100049, China (liuzejian2018@ia.ac.cn, lifanrong2017@ia.ac.cn).}
\thanks{J. Cheng is with the Institute of Automation, Chinese Academy of Sciences,
Beijing 100190, China, the University of Chinese Academy of Sciences,
Beijing 100049, China, and the Center for Excellence in Brain
Science and Intelligence Technology, Chinese Academy of Sciences, Beijing
100190, China (jcheng@nlpr.ia.ac.cn).}}

\maketitle

\begin{abstract}
Deep convolutional neural networks have achieved remarkable progress in recent years. However, the large volume of intermediate results generated during inference poses a significant challenge to the accelerator design for resource-constraint FPGA. Due to the limited on-chip storage, partial results of intermediate layers are frequently transferred back and forth between on-chip memory and off-chip DRAM, leading to a non-negligible increase in latency and energy consumption. In this paper, we propose block convolution, a hardware-friendly, simple, yet efficient convolution operation that can completely avoid the off-chip transfer of intermediate feature maps at run-time. The fundamental idea of block convolution is to eliminate the dependency of feature map tiles in the spatial dimension when spatial tiling is used, which is realized by splitting a feature map into independent blocks so that convolution can be performed separately on individual blocks. We conduct extensive experiments to demonstrate the efficacy of the proposed block convolution on both the algorithm side and the hardware side. Specifically, we evaluate block convolution on 1) VGG-16, ResNet-18, ResNet-50, and MobileNet-V1 for ImageNet classification task; 2) SSD, FPN for COCO object detection task, and 3) VDSR for Set5 single image super-resolution task. Experimental results demonstrate that comparable or higher accuracy can be achieved with block convolution. We also showcase two CNN accelerators via algorithm/hardware co-design based on block convolution on memory-limited FPGAs, and evaluation shows that both accelerators substantially outperform the baseline without off-chip transfer of intermediate feature maps.
\end{abstract}

\begin{IEEEkeywords}
Block Convolution, Memory-Efficient, Off-Chip Transfer, FPGA, CNN Accelerator
\end{IEEEkeywords}

%
\IEEEpeerreviewmaketitle

\section{Introduction}

\IEEEPARstart{I}{n} recent years, deep Convolutional Neural Network (CNN) has become the \textit{de facto} model in many artificial intelligence fields, including computer vision, natural language processing, and robotics. CNN's extraordinary representation ability benefits from the massive training data and the vast computational complexity. High-performance computing devices such as GPUs are commonly used for training deep neural networks. However, as for inference, GPU is no longer an ideal computing platform for deploying CNNs, especially in energy-sensitive edge applications, such as drones. As a result, FPGA-based CNN accelerators have gained increasing attention in recent years in both academia and industry. 

Although FPGA contains a wealth of configurable computing units and memory blocks, there are still considerable challenges when deploying CNN models, especially for low-cost and resource-constraint FPGA. This is mainly due to CNN's three primary features: 1) the vast computing complexity; 2) the huge amount of network weights; 3) the massive intermediate feature maps. Many techniques have been proposed in the algorithm side to reduce the complexity of computation and network weights, such as matrix factorization \cite{matfac}, pruning \cite{NIPS2015_pruning}, low-precision fixed-point quantization \cite{han2015deep}, etc. These methods can effectively alleviate the pressure of on-chip computing and storage. For example, compared with floating-point format, fixed-point operation consumes much less DSP/LUT and on-chip storage. 

However, for efficient handling of the massive intermediate feature maps, there is a lack of research on the hardware side. To accommodate the large volume of feature maps with limited on-chip storage, most accelerators adopt tiling in the spatial and channel dimensions. This introduces a great amount of off-chip data transfers and data processing in the host processor during inference, leading to an increase of latency and energy consumption \cite{qiu2016going}\cite{9408176}. We observe that the conventional spatial tiling is not hardware-friendly. Due to the data dependency at the boundary of tiles, the computing of one tile cannot finish until the boundary pixels of adjacent tiles are available, resulting in a large on-chip memory requirement.

In this paper, we propose block convolution, a memory-efficient alternative of traditional convolution that can completely avoid the off-chip data transfer of intermediate results when deploying large-scale CNN models on memory-limited FPGA. The basic idea of block convolution is to split the feature maps of convolutional layers into independent blocks, thus the calculations of consecutive convolutional layers can be easily fused to maximize the computational density with minimal on-chip buffer requirement. Extensive experiments demonstrate block convolution's broad applicability to various CNN models and the efficiency of accelerator design. The contribution of this paper can be summarized as follows:

\begin{figure}[]
\centering
\subfigure{
\includegraphics[width=0.95\columnwidth]{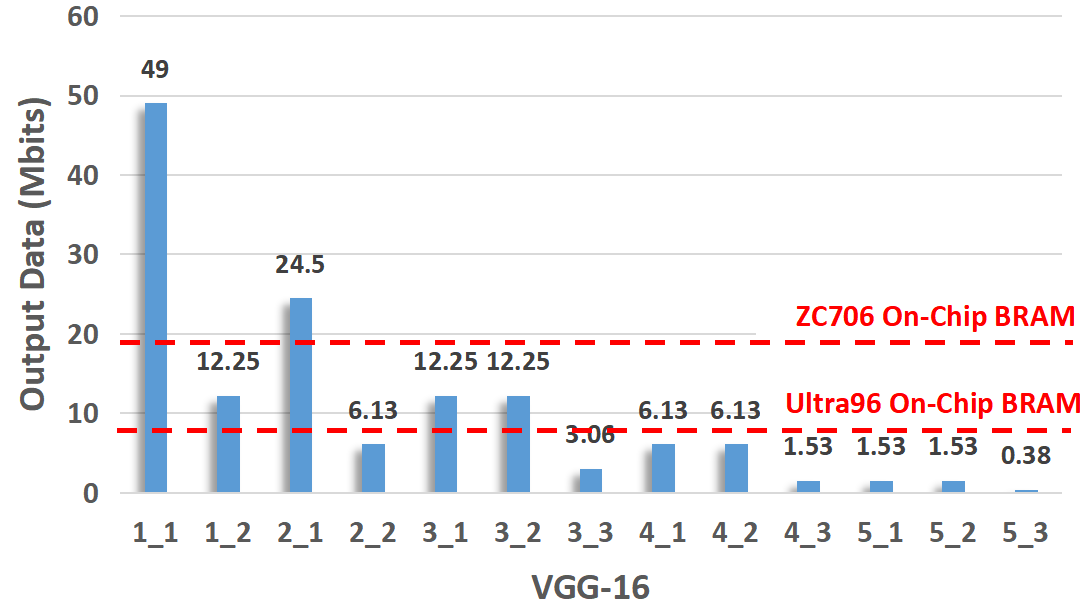}
}
\subfigure{
\includegraphics[width=0.95\columnwidth]{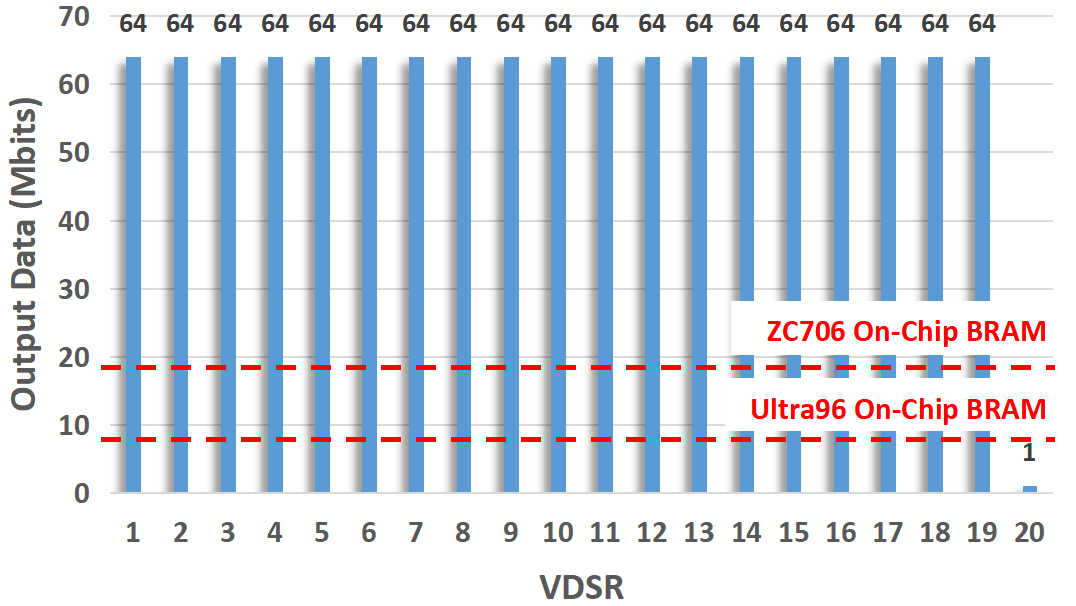}
}
\caption{The volume of intermediate feature maps of VGG-16 and VDSR.}
\label{fig1}
\end{figure}

\begin{itemize}
\item 
We propose a novel, simple, yet efficient operation named block convolution. It is a hardware-friendly alternative to the conventional convolution, which eliminates the tile dependency caused by spatial tiling. With block convolution, multi-layer fusion can be efficiently performed to avoid the off-chip data transfer at run-time when accelerating large-scale CNN on memory-limited FPGA.
\item 
Extensive experiments on ImageNet classification task, COCO object detection task, and Set5 \cite{set5} single image super-resolution task demonstrate that the proposed block convolution can be widely applied to VGG-16 \cite{szegedy2015going}, ResNet-18, ResNet-50 \cite{he2016deep}, MobileNet-V1 \cite{howard2017mobilenets}, SSD \cite{liu2016ssd}, FPN \cite{fpn}, and VDSR \cite{vdsr} with less than 1\% accuracy/mAP/PSNR degradation. In addition, the ablation studies reveal that block convolution has broad adaptability to different blocking patterns, blocking sizes, and can be deployed in combination with fixed-point quantization with negligible accuracy loss.
\item 
    We showcase two CNN accelerators via algorithm/accelerator co-design to evaluate the benefits of block convolution. Specifically, we implement VGG-16 for image classification and VDSR for single image super-resolution on Xilinx ZC706 SoC and Ultra96 MPSoC, respectively. Evaluation results show that both accelerators substantially outperform the baselines in terms of memory efficiency.

\end{itemize}

\section{Block Convolution: Algorithm and Evaluation}

\subsection{Motivation}
\label{motiv}
Unlike ASIC with customizable resources, FPGA consists of a fixed number of on-chip computing and storage resources, which challenges the CNN accelerator design, especially for embedded low-cost FPGA with insufficient on-chip memory. The storage overhead of CNN is mainly caused by network weights and intermediate feature maps. Figure~\ref{fig1} shows the on-chip BRAMs of two FPGAs and the volume of output feature maps of VGG-16 and VDSR for image classification and single image super-resolution, respectively. Their input resolutions are $224\times 224$ and $256\times 256$, and weights and activations are represented as 16-bit fixed-point values. It is obvious that both FPGAs cannot accommodate all the intermediate results of the two models. For example, the output data size of VGG-16's first layer is nearly 50Mbits, which far exceeds the capacity of on-chip memory, and the amount of intermediate results of VDSR is even larger.

To alleviate the pressure of on-chip storage, FPGA-based accelerators often leverage off-chip DRAM to buffer intermediate feature maps. In this scenario, the data transfer size is twice that of the feature maps because the output of the current layer is the input to the next layer. However, frequent data transfer is not efficient. On the one hand, the energy cost of accessing DRAM is several orders of magnitude of SRAM \cite{NIPS2015_pruning}, accounting for a considerable portion of the total energy consumption of an accelerator.
On the other hand, the DRAM bandwidth is shared by several devices in a system, frequent access to DRAM by DNN accelerator will significantly degrade the system's overall performance. 
Therefore, how to efficiently process a large number of intermediate feature maps and access DRAM as little as possible is a critical issue in the design of FPGA-based CNN accelerator.

From Figure~\ref{fig1}, we can notice that the amount of intermediate feature maps in VGG-16 gradually decreases as the network deepens. That is, the first few layers of the network are the storage bottleneck for FPGA deployment. For ZC706, If we can fuse the first four layers and directly calculate the fourth layer's results without caching the whole intermediate layers, the on-chip storage space will be sufficient to accommodate the entire output feature maps, and off-chip data transfer can be eliminated. However, due to the computational dependency between consecutive layers, it is not easy to fuse multiple layers  without buffering large amount of intermediate results, especially when spatial tiling is used. 

\begin{figure}[]
\centering
\subfigure[convolution with overlapped tiling]{
\includegraphics[width=0.9\columnwidth]{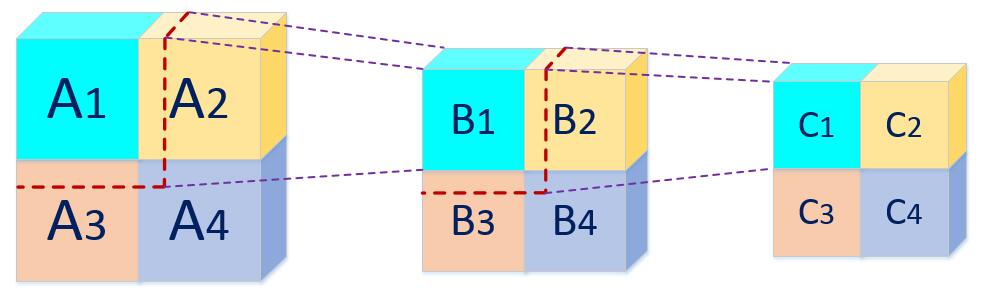}
\label{fig2(a)}
}
\subfigure[block convolution]{
\includegraphics[width=0.9\columnwidth]{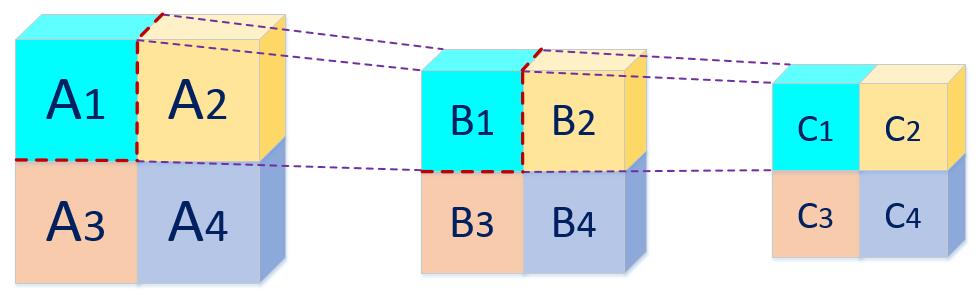}
\label{fig2(b)}
}
\caption{An illustration of conventional convolution with spatial tiling (a) and the proposed block convolution (b).}
\end{figure}

Figure~\ref{fig2(a)} illustrates the data dependency introduced by spatial tiling. There are three consecutive convolutional layers in this example, each layer is partitioned into four overlapped tiles in the spatial dimension. Obviously, The computing of $B_{1}$ cannot finish until all the four tiles of the first layer are available. Similarly, the calculation of $C_{1}$ also depends on all the tiles of the second layer. In other words, given a tile of the first layer, the corresponding output tile in the third layer cannot be directly calculated through layer fusion, leading to a low computing density.
Therefore, reducing the data dependency between spatial tiles is undoubtedly beneficial at both the hardware and software level, regardless of whether the accelerator adopts a multi-core design \cite{shao2019simba} or a single-core design \cite{qiu2016going}.

\begin{figure}[]
\centering
\includegraphics[width=\columnwidth]{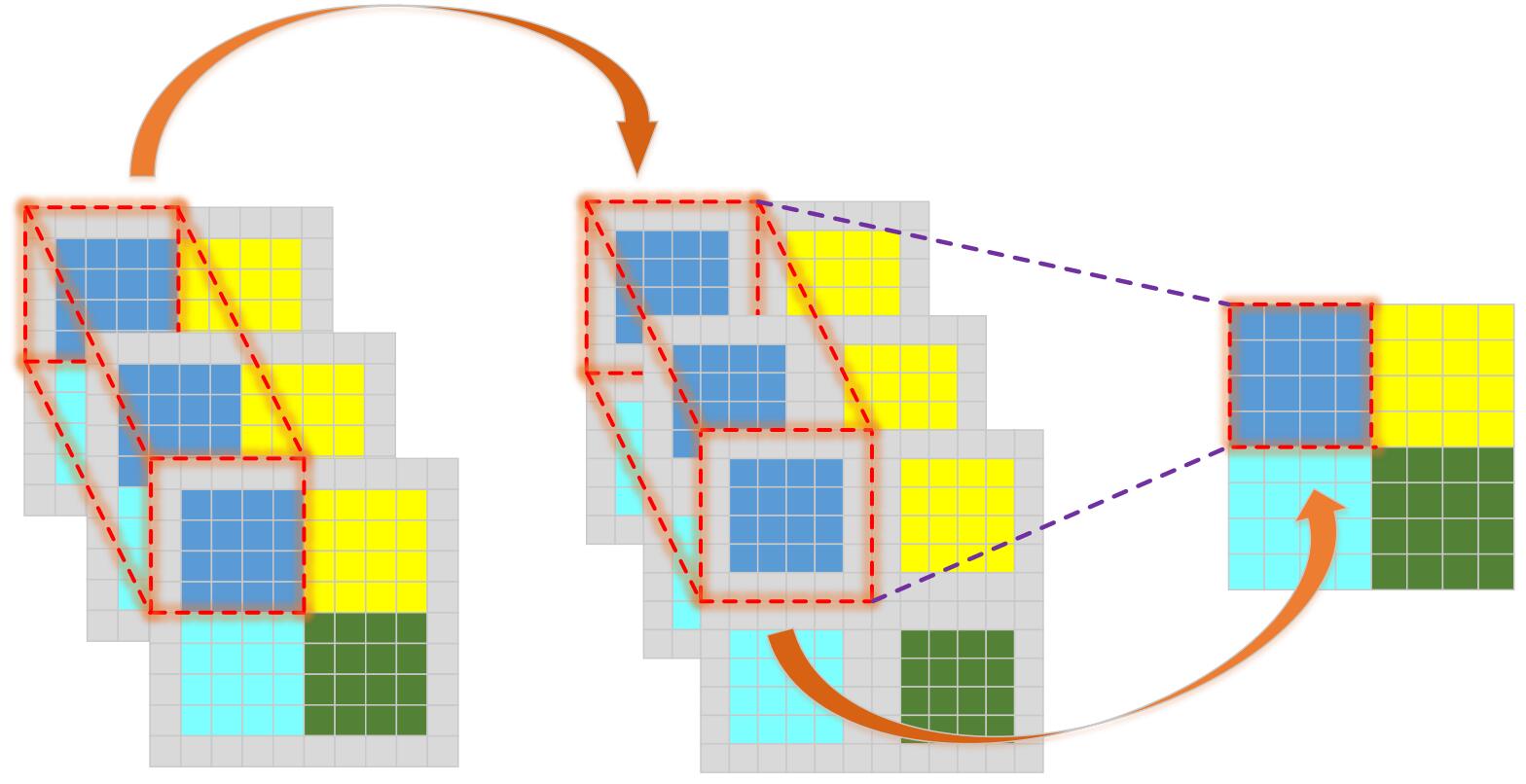}
\caption{An example of block convolution. The input is an $8\times 8\times 3$ tensor, and the filter is a $3\times 3\times 3$ tensor. In original convolution, the number of conv3x3 operation in the spatial dimension is $8\times 8\times 3=192$, whereas in block convolution is $(4\times 4\times 3)\times 4=192$, which remains the same.}
\label{fig3}
\end{figure}

\subsection{Overview}
To address the tile dependency, an intuitive idea is to perform convolution on individual tiles separately in the spatial dimension to improve locality. Hence, we propose block convolution, which is a new operation that restricts convolution within a single tensor block, as shown in Figure~\ref{fig2(b)}. The calculation of each of the four blocks in the second layer merely relies on only one input block in the corresponding position of the first layer. When a block in the second layer is obtained, it can be used immediately to calculate an entire block in the third layer without additional input data. Therefore, with block convolution, the computing of three consecutive convolutional layers can be easily fused without buffering the intermediate feature maps in off-chip memory. Obviously, block convolution is very friendly to the FPGA deployment of large-scale CNNs such as VGG-16 and VDSR in Figure~\ref{fig1}.

\subsection{Algorithm}
\label{algo}
Block convolution employs a \textit{split-concat} computing mechanism, there is no data dependency between the convolution results of adjacent spatial blocks.
For simplicity, we use the example in Figure~\ref{fig3} to illustrate the principle of block convolution. In this example, the input tensor contains 3 feature maps of size $8\times 8$, the size of the convolution filter is $3\times 3\times 3$, and both convolution stride and padding are set to 1. The input feature maps are partitioned into four blocks in the spatial dimension, that is, each block is a 3D tensor of size $4\times 4\times 3$. In the original convolution, for an input feature map of size $[H_{in}, W_{in}]$, the output size can be determined as:
\begin{equation}
\small
H_{out} = \left\lfloor \frac{H_{in} + 2p-k}{s}\right\rfloor+1, W_{out} =\left\lfloor \frac{W_{in} + 2p-k}{s}\right\rfloor+1
\end{equation}
which is $8\times 8$ in this example. 
In block convolution, each input tile of size $5\times 5$ (including one padding) in each feature map is convolved with a $3\times 3$ kernel to obtain a $3\times 3$ output tile, then four $3\times 3$ output tiles are concatenated to form an output feature map of size $6\times 6$, which is inconsistent with the size of the original output feature map. To address this, we conduct \textit{block padding} for each input tile separately before convolution. That is, the $5\times 5$ input tile is symmetrically padded to a $6\times 6$ tile in which the central $4\times 4$ 
pixels are from the original feature map, as depicted in Figure~\ref{fig3}. Through this way, four $4\times 4$ output tiles can be calculated independently and concatenated to form an $8\times 8$ output feature map. 

More generally, converting conventional convolution with stride $s$, padding size $p$, and kernel size $k$ into block convolution is equivalent to find the proper blocking number $N$ and block padding size $p_{t}$ such that:
\begin{equation}
\small
\left\lfloor \frac{I + 2p-k}{s}\right\rfloor+1=N(\left\lfloor \frac{I/N + 2p_{t}-k}{s}\right\rfloor+1)
\label{general eq}
\end{equation}
where $I$ denotes the input feature maps of size $[H_{in}, W_{in}]$.
It can be seen that block convolution essentially splits an entire feature map into several sub-feature maps in the spatial dimension. Then convolution is performed on each sub-feature map independently, and finally the results are spliced together. When the kernel size is 1, the block convolution is exactly the pointwise ($1\times 1$) convolution. 

Note that the computational complexity (FLOPs) of block convolution is completely the same as the original convolution, which is different from the model compression techniques that aim at reducing computations or network parameters.

\begin{figure}[]
\centering
\includegraphics[width=\columnwidth]{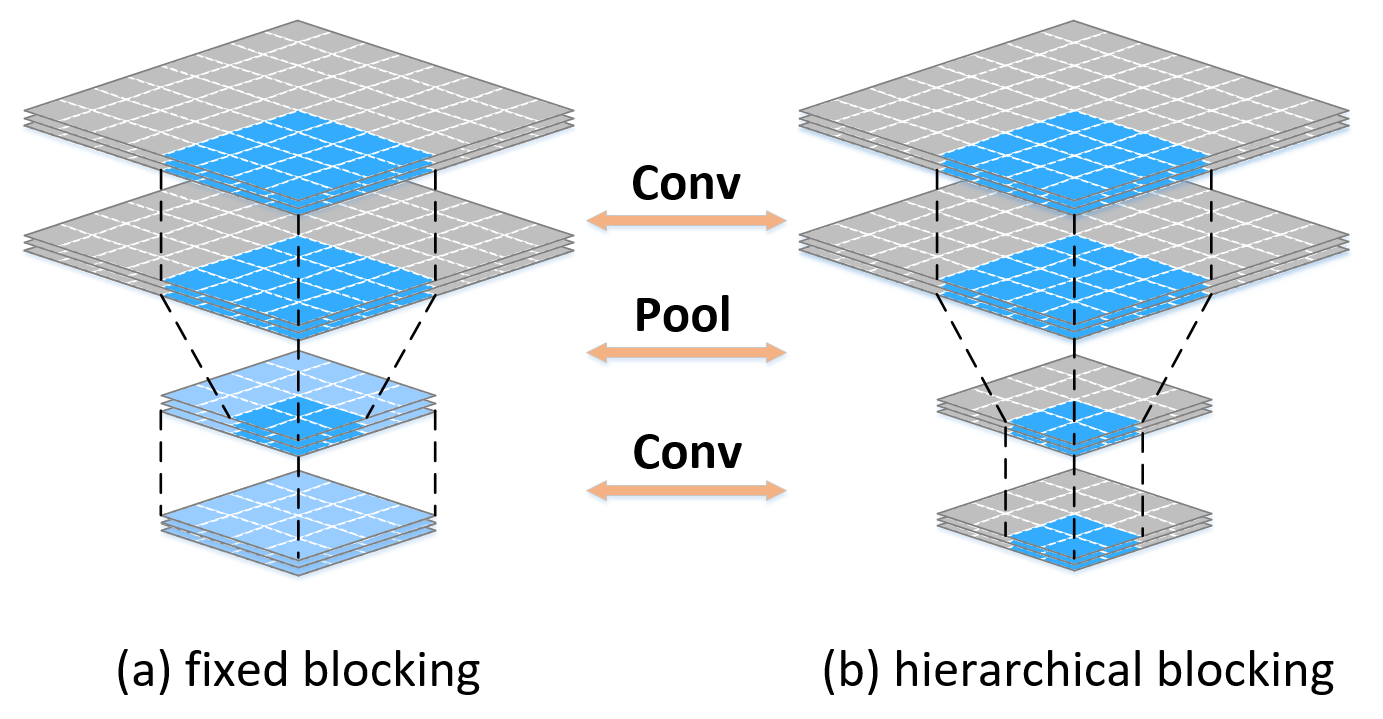}
\caption{An illustration of blocking patterns: fixed blocking (a) and hierarchical blocking (b). The blue region in each layer denotes a block.}
\label{fig4}
\end{figure}

\begin{table*}[]
\centering
\caption{Top-1 accuracy on ImageNet classification task. We list the accuracies from both the official torchvision library and our trained baseline in which we replace the original stride operation (stride$>1$) to max-pooling.}
\resizebox{\textwidth}{!}{
\begin{tabular}{c|c|c|c|c|c}
\hline
\textbf{}                                  & Torchvision \cite{accuracy} & Baseline & BConv + Training from Scratch & BConv + Fine-tuning  & Blocking Ratio \\ \hline
VGG-16                           & $71.59\%$       & $71.59\%$           & $70.47\% \quad \downarrow1.12\%$                                             & $71.45\% \quad \downarrow0.14\%$                                  &   $76.92\%$                      \\ 
ResNet-18                         & $69.76\%$           & $70.60\%$           & $69.94\% \quad \downarrow0.66\%$                                            & $71.21\% \quad \uparrow\textbf{0.61\%}$                                  &  $76.47\%$                       \\ 
ResNet-50                         & $76.15\%$           & $75.86\%$           & $75.42\% \quad \downarrow0.44\%$                                            & $76.67\% \quad  \uparrow \textbf{0.81\%}$                                  &   $81.63\%$                      \\ 
\multicolumn{1}{l|}{MobileNet-V1} & $70.60\%$           & $72.29\%$           & $72.05\% \quad \downarrow0.24\%$                                            & $71.76\% \quad \downarrow0.53\%$                                  &   $44.44\%$                      \\ \hline
\end{tabular}
}
\label{tab1}
\end{table*}

\subsection{Blocking Pattern}
According to the characteristics of CNN's architecture, we propose two blocking patterns for multi-layer fusion: \textit{fixed blocking} and \textit{hierarchical blocking}.

{\bf Fixed Blocking.} 
As depicted in Figure~\ref{fig4}(a), fixed blocking keeps the block size consistent through layers. After pooling, adjacent output blocks with reduced resolution are spliced into a large block for subsequent processing. It can be seen that as the network deepens, the number of blocks in a layer will decrease, and the receptive field of output blocks will increase.

{\bf Hierarchical Blocking.}
Unlike fixed blocking, the number of blocks is constant in each layer under hierarchical blocking, as illustrated in Figure~\ref{fig4}(b). As the network deepens, the size of individual block gradually becomes smaller. Compared with fixed blocking, the block's receptive field in each layer stays unchanged under hierarchical blocking. In other words, the entire network is divided into several independent sub-networks along the spatial dimension, except for the fully-connected layers.

\subsection{Skip Connection and Depthwise Convolution}
Block convolution can be applied to various convolutional architectures, including the plane structure that stacks several convolutional layers in a straightforward manner, or the residual structure that contains skip connection. The main operations in these structures include $k\times k~(k>1)$ convolution, pointwise convolution, downsampling, and element-wise summation, in which the latter three operations are naturally splittable in the spatial dimension. Besides, block convolution can also be applied to depthwise convolution, a commonly used operation in compact models such as MobileNets. We will show the evaluation results in the next section.

\subsection{Experiments}
\label{experiment}
{\bf Methodology.}
To demonstrate the effectiveness of block convolution, we conduct extensive experiments on three computer vision tasks: image classification, object detection, and single image super-resolution. Specifically, we investigate the impact of blocking size, blocking pattern, block padding, and fixed-point quantization on accuracy for various networks with block convolution.
For a fair comparison, we keep the hyperparameters during network training/fine-tuning consistent with the original network, including data augmentation, number of iterations, learning rate, weight decay, optimization method, etc., and do not use any additional tricks. All the experiments are conducted using PyTorch framework. Code is available at \url{https://github.com/zejiangp/BlockConv}.

{\bf ImageNet Classification.}
In this experiment, we benchmark block convolution on four deep networks: VGG-16, ResNet-18, ResNet-50, and MobileNet-V1. 
Table~\ref{tab1} lists the top-1 classification accuracy of the four networks and their block convolution variants on the ILSVRC-12 dataset. From Equation~\ref{general eq}, we can notice that the block padding can be asymmetric, especially when convolutional stride is larger that 1. For simplicity, we modify the convolutional layers with stride $s$ to those with stride 1 followed by an $s\times s~(s>1)$ max pooling layer. Then we train these modified networks from scratch with the same hyperparameters, and serve them as our baselines. Surprisingly, from the results in the first two columns in Table~\ref{tab1} , we can see that three out of four networks maintain or improve the accuracy after modification, resulting in a stronger baseline.
In this experiment, we adopt fixed blocking of size $28\times 28$ with zero block padding, and block all the convolutional layers (including input layer) whose resolution are larger than $28\times 28$. The last column in Table~\ref{tab1} shows the proportion of blocked layers to the total convolutional layers for each network.

For block convolution, we evaluate two training strategies: training from scratch and fine-tuning from the pre-trained model. From the results in Table~\ref{tab1}, it can be seen that the accuracies of all the eight blocked networks obtained by both strategies are very close to the baselines. VGG-16, ResNet-18, and ResNet-50 achieve higher accuracies through fine-tuning while MobileNet-V1 prefers training from scratch.
Compared with the baseline models, the accuracy degradation of all the blocked networks except for the trained VGG-16 is no more than 1\%.
In particular, the accuracies of the fine-tuned ResNet-18 and ResNet-50 with block convolution are improved by 0.61\% and 0.81\%, respectively. 
These results indicate that although block convolution introduces computational bias, there is little difference between block convolution and conventional convolution in terms of classification accuracy.

\begin{figure}[]
\centering
\includegraphics[width=\columnwidth]{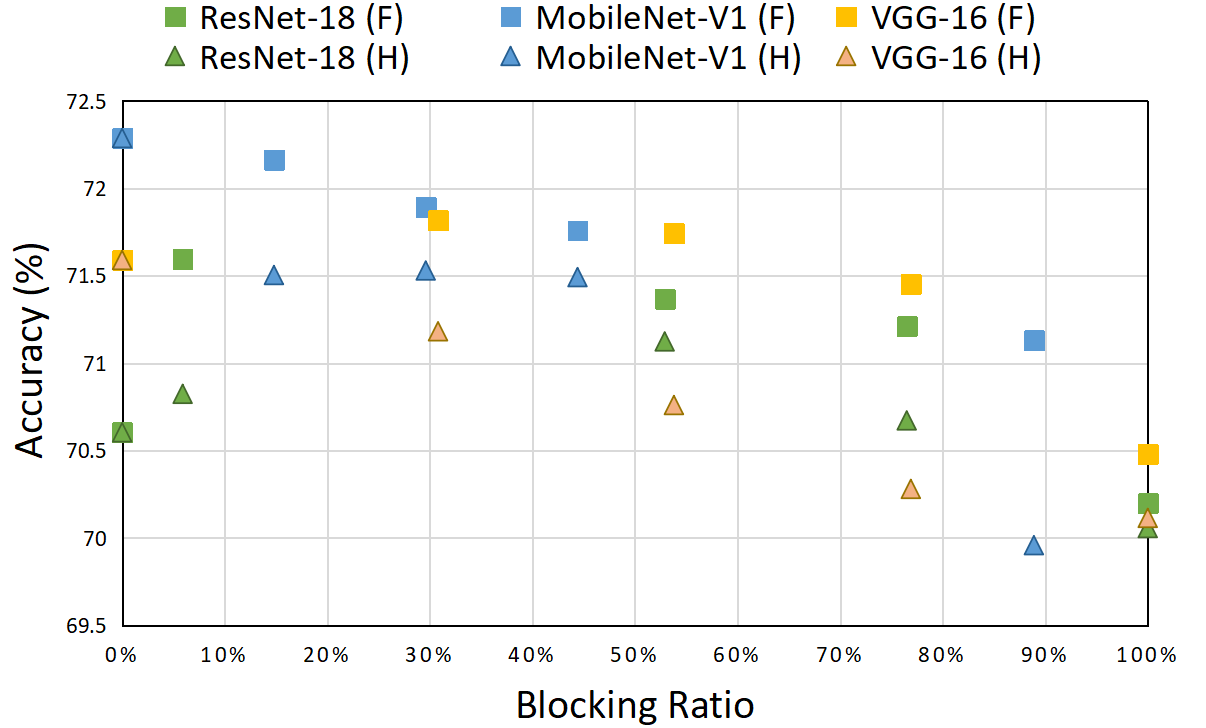}
\caption{Top-1 accuracy of blocked networks with respect to blocking ratio under fixed blocking (F) and hierarchical blocking (H).}
\label{fig5}
\end{figure}

As for the impact of blocking patterns on classification accuracy, we evaluate both fixed blocking and hierarchical blocking on ResNet-18, MobileNet-V1, and VGG-16. Figure~\ref{fig5} shows the accuracies of all the blocked networks with respect to blocking ratio, in which $H_{i\times i}$ denotes that each convolutional layer is partitioned into $i\times i$ independent blocks through hierarchical blocking, 
and $F_{i}$ stands for the block size in each convolutional layer is fixed to $i\times i$ under fixed blocking. As before, we block the convolutional layers as many as possible, including the input layer. Zero block padding is also used in this experiment.
From the results in Figure~\ref{fig5}, we can conclude that: 
\begin{enumerate}
    \item \textit{As the number of blocked layers increases, the accuracy decreases for both fixed blocking and hierarchical blocking in most cases.} This is reasonable  because the more convolutional layers are blocked, the less information fusion before the classification layer, which will hurt the representation ability of the network;
    \item \textit{Fixed blocking consistently outperforms hierarchical blocking under the same blocking ratio.} This can be attributed to two points. First, in contrast to hierarchical blocking, there is information fusion between independent blocks under fixed blocking. This helps expand the receptive field of a block and encourages the transmission of information across different blocks, which is beneficial to maintaining accuracy. Second, with the same blocking ratio, the number of blocks in each layer under hierarchical blocking is no less than that under fixed blocking, which inevitably introduces larger accuracy degradation;
\end{enumerate}

\begin{table}[]
\centering
\caption{Accuracy of non-square blocking on ResNet-18.}
\label{tab2}
\resizebox{0.5\columnwidth}{!}{
\begin{tabular}{c|c}
\hline
\multicolumn{1}{c|}{} & \multicolumn{1}{c}{Accuracy} \\ \hline
Baseline              & $70.60\%$                                          \\ 
$F_{28\times 56}$              & $71.31\%\quad\uparrow0.71\%$                    \\ 
$H_{4\times 1}$                & $71.14\%\quad\uparrow0.54\%$                     \\ 
$H_{1\times 4}$                & $71.22\%\quad\uparrow0.62\%$                     \\ \hline
\end{tabular}
}
\end{table}

In the above experiments, only square blocking is considered. We also evaluate three rectangular blocking shapes on ResNet-18: $F_{28\times 56}$, $H_{1\times 4}$, and $H_{4\times 1}$. Results are shown in Table~\ref{tab2}. We can see that the accuracies of ResNet-18 under all the three non-square blocking configurations are higher than the non-blocked baseline. 

\begin{figure}[]
\centerline{\includegraphics[width=0.9\columnwidth]{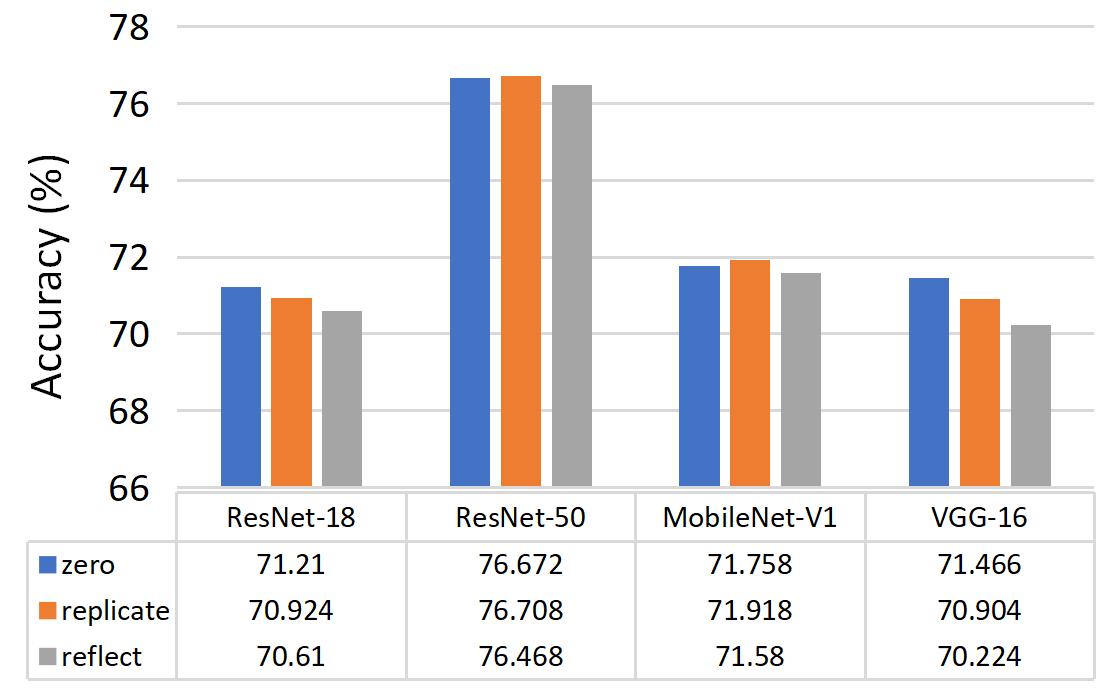}}
\caption{Impact of block padding on classification accuracy.}
\label{fig6}
\end{figure}

As mentioned in Section~\ref{algo}, the function of block padding is to keep the output feature map consistent in size with the original one. Here we investigate three different padding methods: zero padding, replicate padding, and reflect padding. Among them, replicate padding copies the boundary pixels outwards, while reflect padding symmetrically pads values with the boundary as the axis. Figure~\ref{fig6} shows the accuracies of four networks with $28\times 28$ fixed blocking. It can be seen that for ResNet-18 and VGG-16, zero padding can achieve the highest accuracy, whereas replicate padding is more preferable for ResNet-50 and MobileNet-V1. As for the hardware implementation of block padding, it is not necessary to explicitly pad each tiles in data buffer.
Since the computing of each tile only relies on itself, in all the three padding cases, block padding can be merged into convolution process either by on-power initialization or
on-the-fly manipulating of memory address, which leads to negligible overhead in terms of control and computing.

\begin{figure}[]
\centerline{\includegraphics[width=0.9\columnwidth]{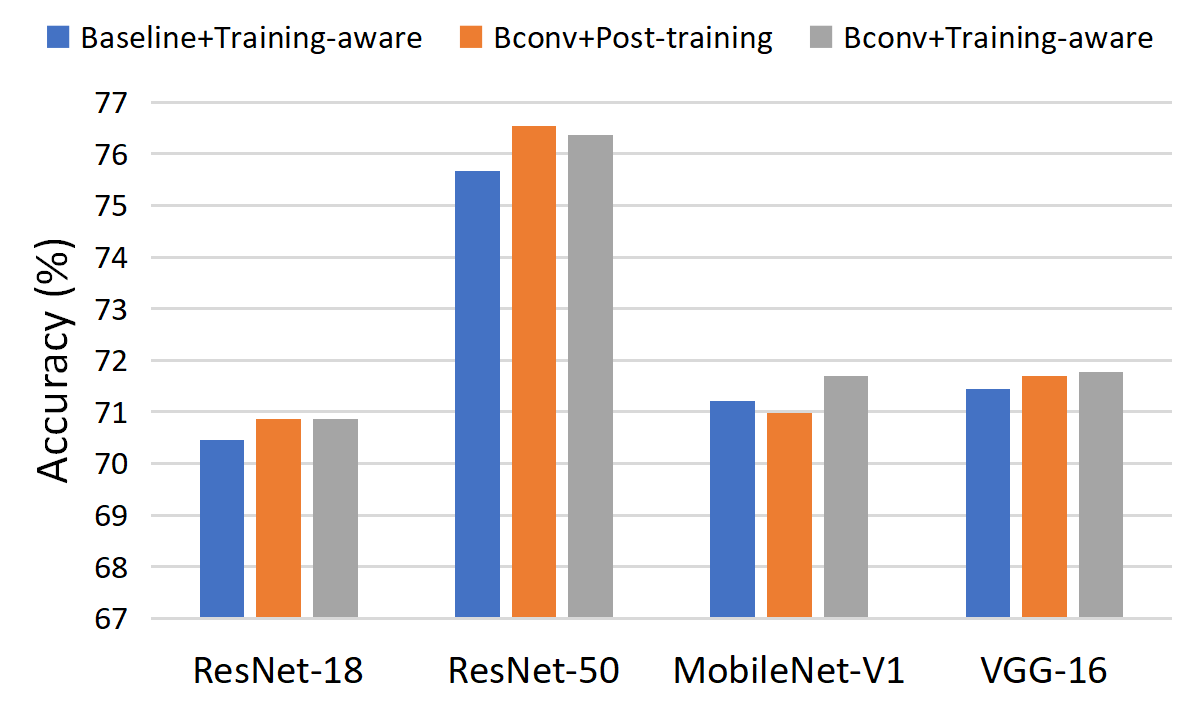}}
\caption{Results of 8-bit quantization for block convolution.}
\label{fig7}
\end{figure}

Finally, we evaluate the impact of 8-bit fixed-point quantization on block convolution. Figure~\ref{fig7} shows the quantization results for four baseline networks and their $F_{28}$ blocked variants. We investigate both training-aware quantization and post-training quantization on blocked networks. All the experiments are conducted using the Neural Network Distiller \cite{nzmora2019distiller}.
It is clear to see that with training-aware quantization, all the 8-bit quantized blocked networks outperform their non-blocked counterparts. Even with post-training quantization, three out of four blocked networks can achieve higher accuracy than the baseline networks.

\begin{table}[]
\centering
\caption{Benchmark for object detection.}
\label{tab3}
\resizebox{0.35\textwidth}{!}{%
\begin{tabular}{l|cc}
\hline
    & \multicolumn{1}{c}{Backbone Network} & \multicolumn{1}{c}{Input Size} \\ \hline
SSD & VGG-16                               & $300\times 300$                        \\ \hline
FPN & ResNet-50                            & $1333\times 800$                       \\ \hline
\end{tabular}%
}
\end{table}

\begin{table*}[]
\centering
\caption{PSNR of VDSR and its block convolution variants on Set5 dataset.}
\label{tab4}
\resizebox{\textwidth}{!}{%
\begin{tabular}{c|ccccc}
\hline
Scale &
  VDSR &
  \begin{tabular}[c]{@{}c@{}}VDSR+BConv \\ ($H_{2\times 2}$)\end{tabular} &
  \begin{tabular}[c]{@{}c@{}}VDSR+BConv\\ (Fixed)\end{tabular} &
  \begin{tabular}[c]{@{}c@{}}VDSR+BConv\\  ($H_{2\times 2}$, Blocking Depth=2)\end{tabular} &
  \begin{tabular}[c]{@{}c@{}}VDSR+BConv\\ ($H_{2\times 2}$, Blocking Depth=4)\end{tabular} \\ \hline
$\times$2 & 37.53 & 37.07 & 37.23 & 37.44 & 37.34 \\
$\times$3 & 33.69 & 33.4  & 33.44 & 33.65 & 33.55 \\
$\times$4 & 31.34 & 31.01 & 31.07 & 31.31 & 31.21 \\ \hline
\end{tabular}%
}
\end{table*}

\begin{figure}[]
\centerline{\includegraphics[width=0.9\columnwidth]{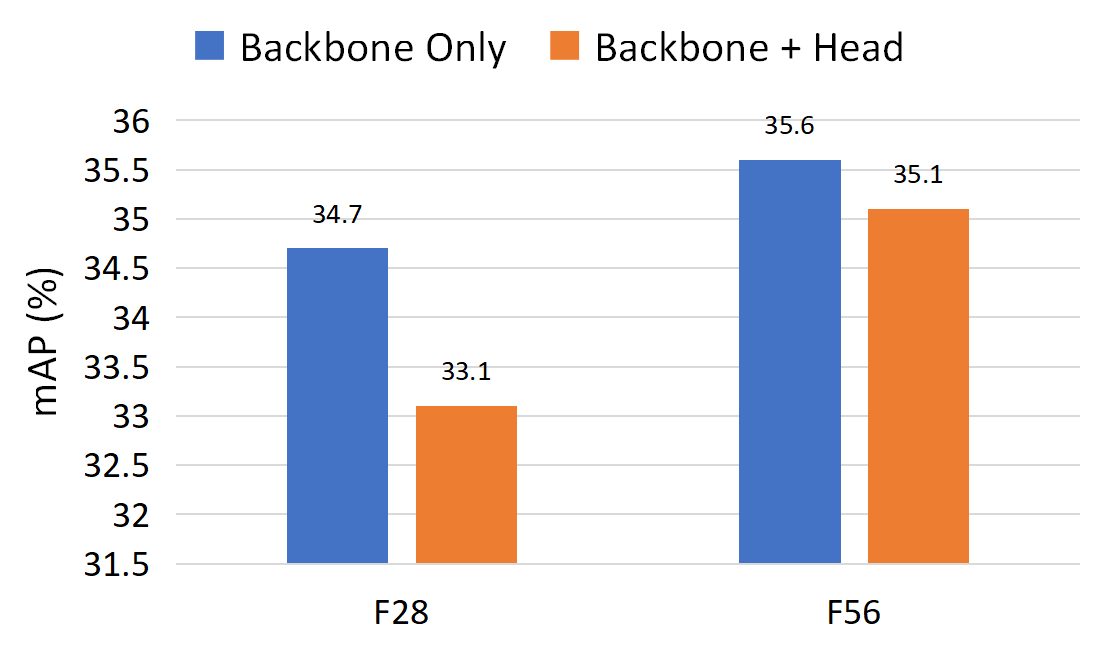}}
\caption{Result of FPN on COCO dataset.}
\label{fig8}
\end{figure}

{\bf COCO Object Detection.}
In this experiment, we choose SSD and FPN for evaluation. The model configuration is shown in Table~\ref{tab3}. During training, we use the pre-trained $F_{28}$ VGG-16 and ResNet-50 to initialize the backbone's parameters of the two detection models, and the parameters of all the detection heads are randomly initialized. As before, we maximize the number of blocked layers of the backbone networks. Results are shown in Table~\ref{tab5}. When block convolution is applied, the mAP drop of FPN and SSD are only 1\% and 1.8\%, respectively. Note that for $F_{28}$, the blocking ratios in these two backbone networks are more than 75\%. If the blocking size is enlarged, such as $56\times 56$, the mAP degradation can be alleviated. This can be concluded from Figure~\ref{fig8}, in which the $F_{56}$ FPN consistently outperform $F_{28}$ with only 0.1\% mAP loss compared with the non-blocked baseline.

\begin{table}[]
\centering
\caption{Results of two object detection models on COCO dataset.}
\label{tab5}
\resizebox{0.48\textwidth}{!}{%
\begin{tabular}{c|cccccc}
\hline
          & AP   & AP@0.5 & AP@0.75 & AP$_{s}$ & AP$_{m}$ & AP$_{l}$ \\ \hline
FPN       & 35.7 & 55.8   & 38.3   & 20.2  & 39.5  & 46.2  \\
FPN+Bconv & 34.7 & 55     & 37     & 20.5  & 38.3  & 45.2  \\ \hline
SSD       & 25.4 & 43.5   & 26.1   & 6.9   & 27.7  & 42.7  \\
SSD+Bconv & 23.6 & 41     & 24.1   & 5.3   & 25    & 40.3  \\ \hline
\end{tabular}%
}
\end{table}

Since the detection heads also contain convolutional layers, we further investigate the impact of block convolution on the detection heads for FPN. As shown in Figure~\ref{fig8}, we compare two settings: 1) only backbone network is blocked; 2) both backbone network and detection heads are blocked. It can be seen that for $F_{28}$ and $F_{56}$, blocking detection heads introduces an additional 1.6\% and 0.5\% mAP loss, respectively. 

It is worth noting that although the detection head is more sensitive to block convolution, the resolution of the detection heads is much smaller than the input resolution and will not cause a serious storage bottleneck. Therefore, in the FPGA-based accelerator design, blocking the detection heads is not always necessary, thus the accuracy can be easily maintained.

{\bf Single Image Super-Resolution.}
CNN-based single image super-resolution is challenging for hardware acceleration due to the high computational complexity and storage overhead. First, although the image size of the training dataset is relatively small (such as $41\times 41$ in Set5 \cite{set5}), the input image in real-world application is very large, such as $1920\times 1080$, which makes the amount of calculation several times greater than that of small input. Second, 
unlike image classification and object detection, the networks commonly used for single image super-resolution do not have the property of reducing the feature map's resolution as the network goes deepen. For example, in VDSR \cite{vdsr}, the resolution of all the intermediate layers is exactly the same as the input. This means that each layer will produce a vast volume of intermediate results during inference, leading to a significant challenge to efficient deployment on memory-constraint FPGA.

In this experiment, we verify the effectiveness of block convolution on VDSR. During training, we replace the original convolution with block convolution (with zero padding) and keep the hyperparameters consistent with the original model.
First of all, we partition the feature maps of size $41\times 41$ in each layer into four blocks of the same size and train from scratch. Results are shown in the first two columns in Table~\ref{tab4}, it can be seen that the PSNR loss is no more than 0.5 for all the three scale factors.
Next, we employ fixed blocking to partition the feature maps of size $41\times 41$ to four parts: $28\times 28$, $28\times 13$, $13\times 28$, and $13\times 13$. Result is shown in the third column. Although this blocking pattern introduces irregularity, we can notice that the PSNR is surprisingly higher than that of $2\times 2$ hierarchical blocking.

In the above experiment, we block all the layers in the VDSR network, which makes it possible to execute inference in sequence on memory-limited FPGA without the aid of off-chip buffers. To explore the trade-off between accuracy and the volume of off-chip data transfer, we conduct an experiment to examine the blocking depth. We block every $n$ consecutive layer followed by a normal convolutional layer, i.e., the blocking depth is $n$. In this scenario, information fusion and off-chip transfer occur at each normal convolutional layer. We set $n=2$ and 4 in this experiment. Results are shown in the last two columns in Table~\ref{tab4}.
It can be seen that information fusion in the intermediate layers is helpful to improve PSNR, and the more information fusion the closer the PSNR is to the original model.

\section{FPGA-based CNN Accelerator Design and Evaluation}
\label{accelerator}

In this section, we demonstrate the effectiveness and efficiency of block convolution on FPGA-based CNN accelerator design. We first briefly overview the design principles of multiple-layer fusion, then showcase two accelerators for large-scale VGG-16 and VDSR, respectively.

\begin{figure*}
\subfigure[MobileNet-V1]{
  \centering
  \includegraphics[width=0.45\textwidth]{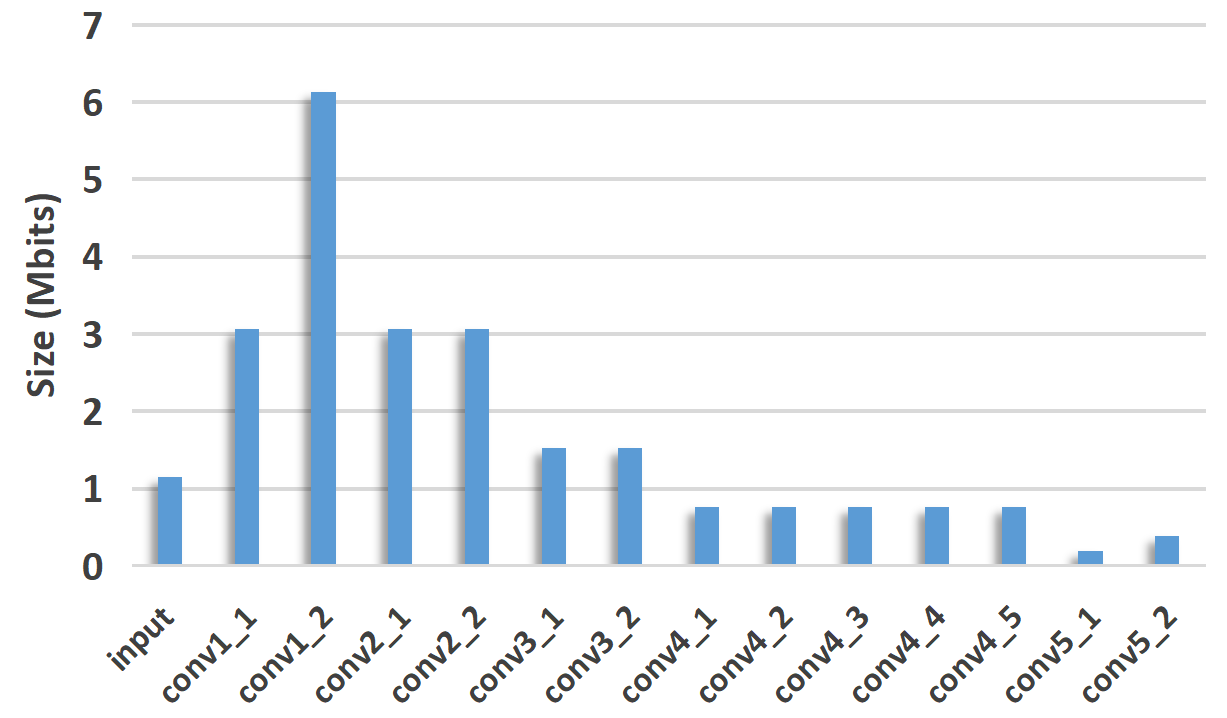}  
}
\subfigure[ResNet-18]{
  \centering
  \includegraphics[width=.5\textwidth]{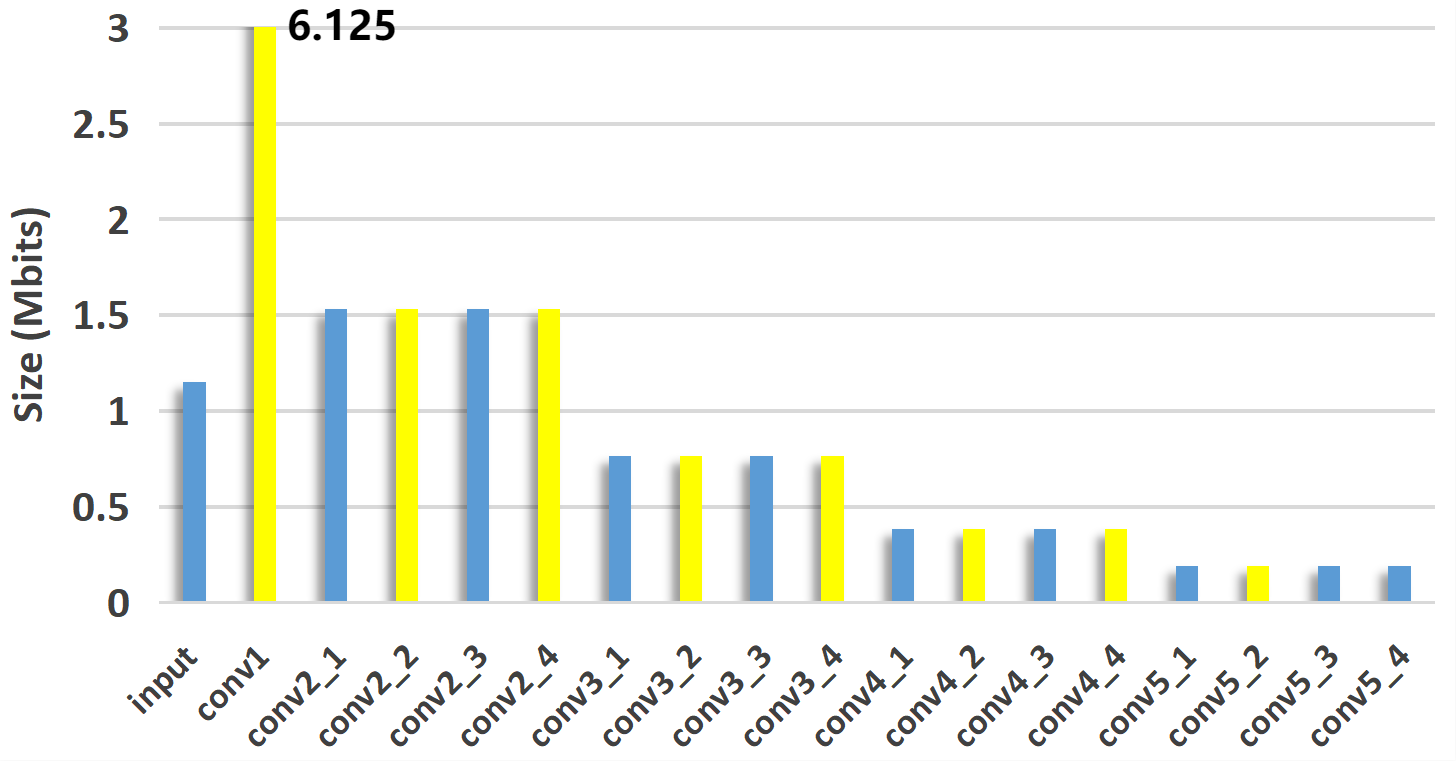}  
}
\newline
\subfigure[ResNet-50]{
  \centering
  \includegraphics[width=0.95\textwidth]{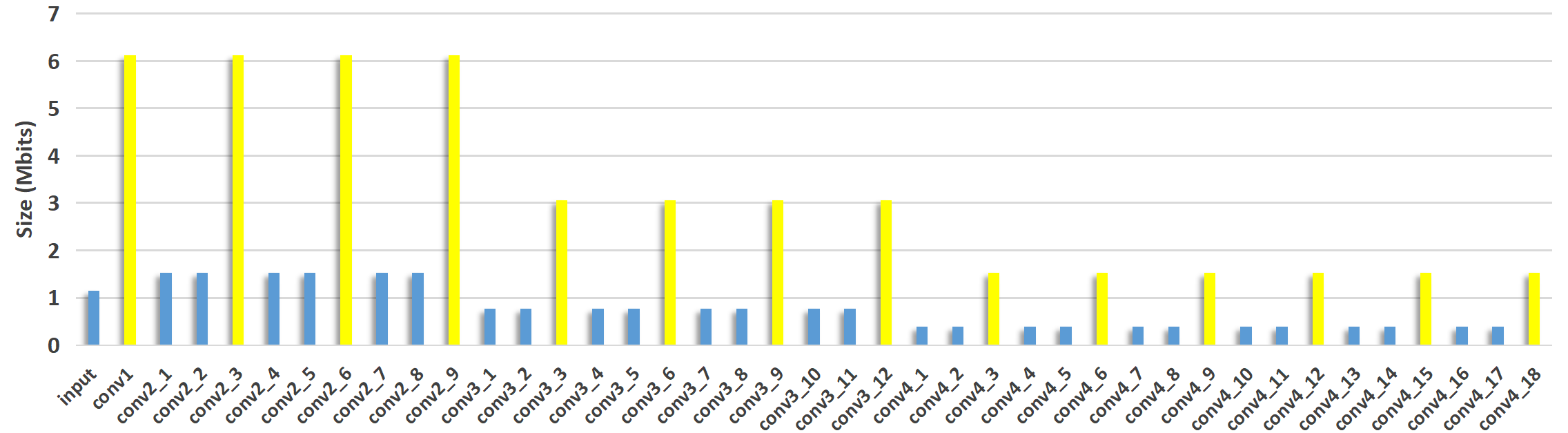}  
}
\caption{Visualization of feature map size (Mbits) of convolutional layers. The layers marked yellow refer to the first layer of a residual block. We omit the convolutional layers after conv4 in ResNet-50.}
\label{fig9}
\end{figure*}

\subsection{Overview}
The inference of a network is often performed layer by layer in a typical CNN accelerator due to the on-chip resource constraint. The main drawback of this strategy lies in the heavy off-chip traffic which is inefficient in terms of latency and energy consumption. However, with block convolution, the massive off-chip transfer of intermediate results can be significantly reduced or eliminated through efficient multi-layer fusion.

\begin{lstlisting}[frame=lines,caption={Pseudo code of the baseline accelerator in \cite{qiu2016going}. $T_{r},T_{c},T_{m},T_{n}$ is the tiling size of height, width, output channel, and input channel, respectively. $N_{pe}$ is the number of parallel PEs. For simplicity, we integrate bias term, non-linear transformation and pooling into a single function named Norm.},captionpos=b,basicstyle=\scriptsize,label={listing1},language={[ANSI]C}]
for(r=0; r<R; r+=Tr){
  for(c=0; c<C; c+=Tc){
    for(m=0; m<M; m+=Tm){
      for(n=0; n<N; n+=Tn){
        //read input data and parameters from DDR
        Read(In,W,r,c,m,n);
        for(tm=0; tm<Tm; tm+=Npe){ #UNROLL
          for(tr=0; tr<Tr; ++tr){
            for(tc=0; tc<Tc; ++tc){
              for(tn=0; tn<Tn; ++tn){ #UNROLL
                for(i=0; i<K; ++i){ #UNROLL
                  for(j=0; j<K; ++j){ #UNROLL
                    Out[tm][r][c]+=
                    W[tm][tn][i][j]*In[tn][S*r+i][S*c+j];
              } } }
              Norm(Out,tm,tr,tc);
        } } }
        //write intermediate data to DDR
        Write(Out,r,c,m,n);
} } } }
\end{lstlisting}

A key to implementing block convolution is determining the fusion depth for a given on-chip memory budget. To avoid off-chip transfer, we can simply fuse multiple layers until a layer's entire output feature maps can be accommodated on-chip. This is often possible since the feature map size will decrease as the network deepens. Figure~\ref{fig9} visualizes the feature map size of convolutional layers of ResNet-18, ResNet-50, and MobileNet-V1 with a $224\times 224$ input resolution. Suppose that we have a memory budget of 7.6Mb, which is the on-chip BRAM capacity of the low-end ZU3EG FPGA in Xilinx Ultra96 MPSoC. For MobileNet-V1, layer conv1\_2 is the main bottleneck. However, with block convolution, we can simply fuse the first four layers and store the output of layer conv2\_1 on-chip. Similar layer fusion strategy can be applied to ResNet-50 and ResNet-18. A slight difference is that an additional buffer is needed for storing a copy of the input tile of a residual block.

\begin{figure*}
\centering
\subfigure[an example of multi-layer fusion]{
\begin{minipage}[b]{0.35\textwidth}
\centering
\includegraphics[width=\textwidth]{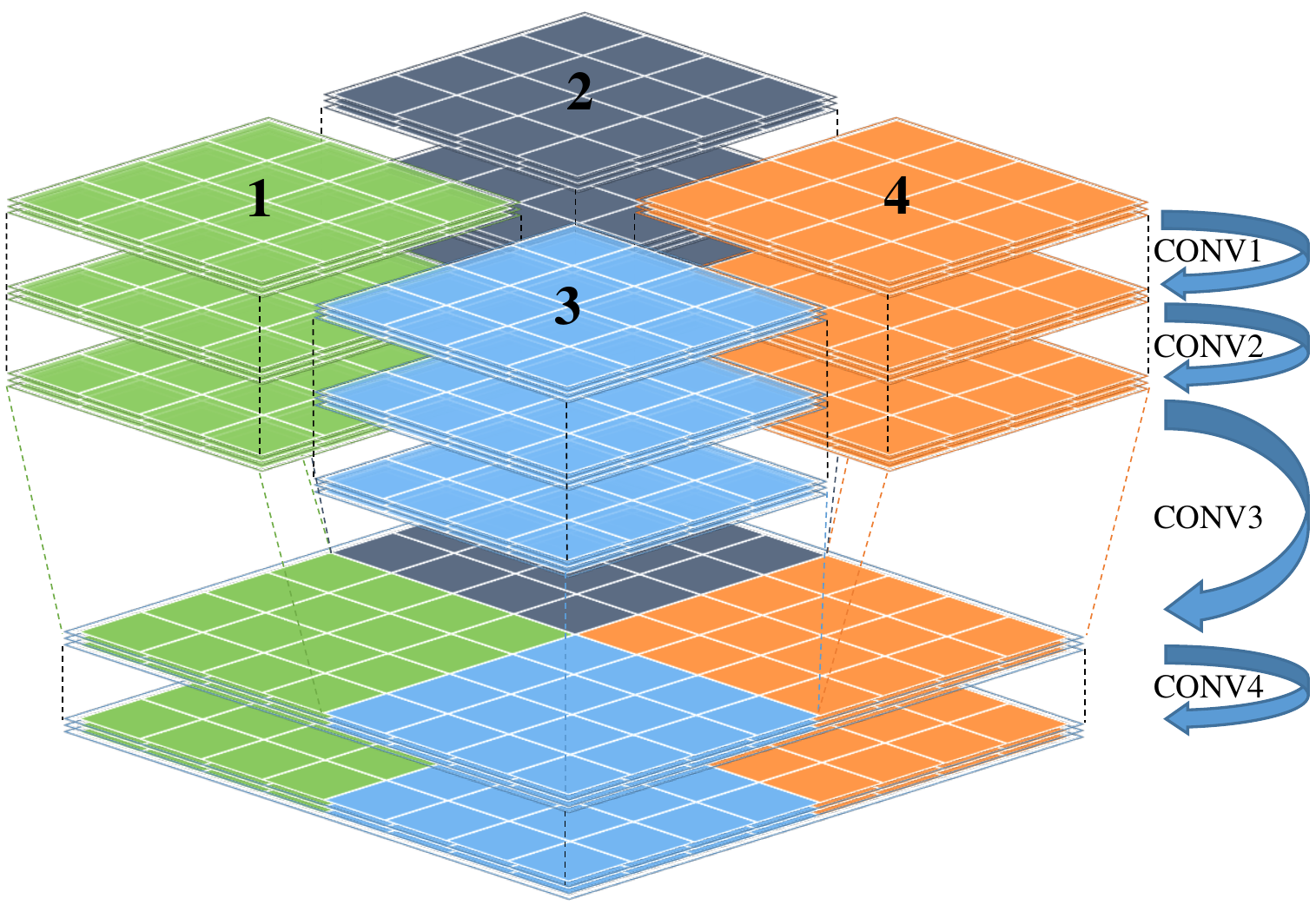}
\end{minipage}}
\subfigure[CONV1 for block 1]{
\begin{minipage}[b]{0.3\textwidth}
\centering
\includegraphics[width=\textwidth]{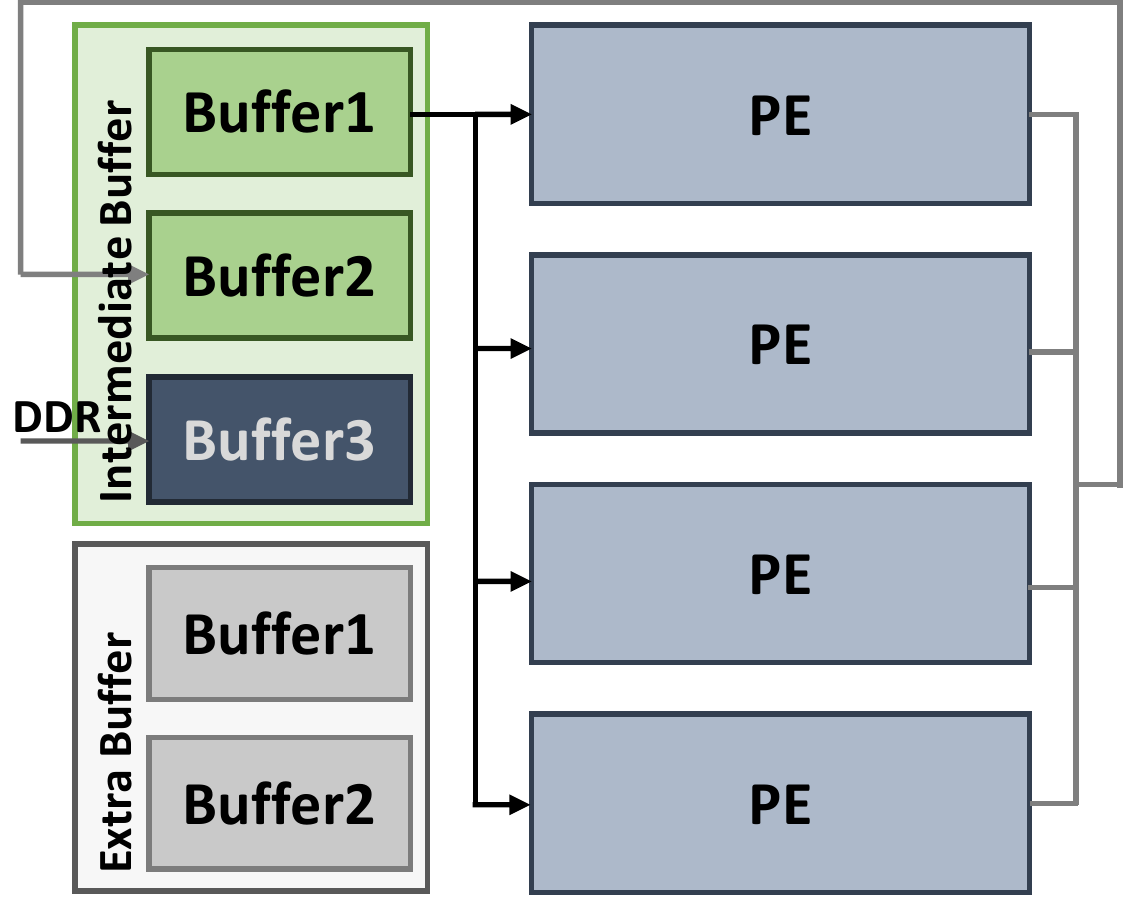}
\end{minipage}}
\subfigure[CONV2 for block 1]{
\begin{minipage}[b]{0.3\textwidth}
\centering
\includegraphics[width=\textwidth]{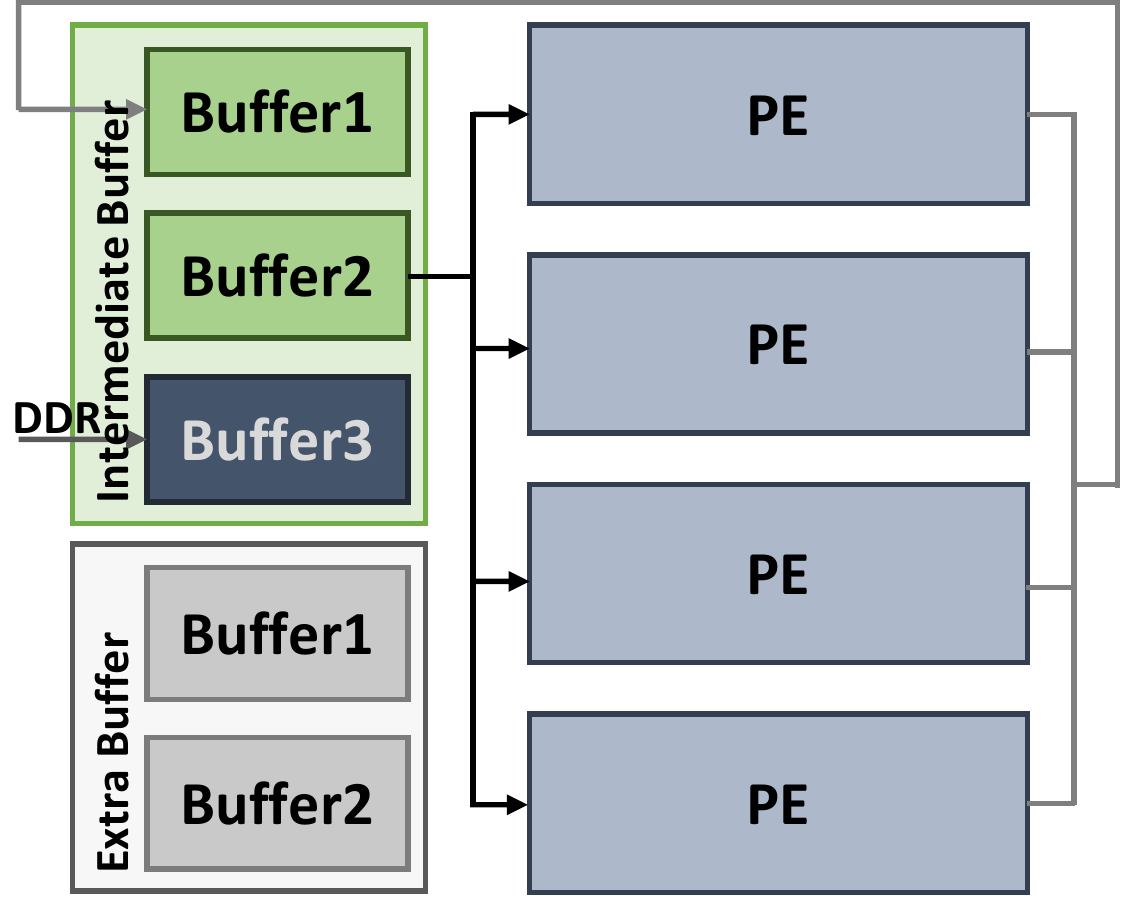}
\end{minipage}}
\subfigure[CONV3 for block 1]{
\begin{minipage}[b]{0.3\textwidth}
\centering
\includegraphics[width=\textwidth]{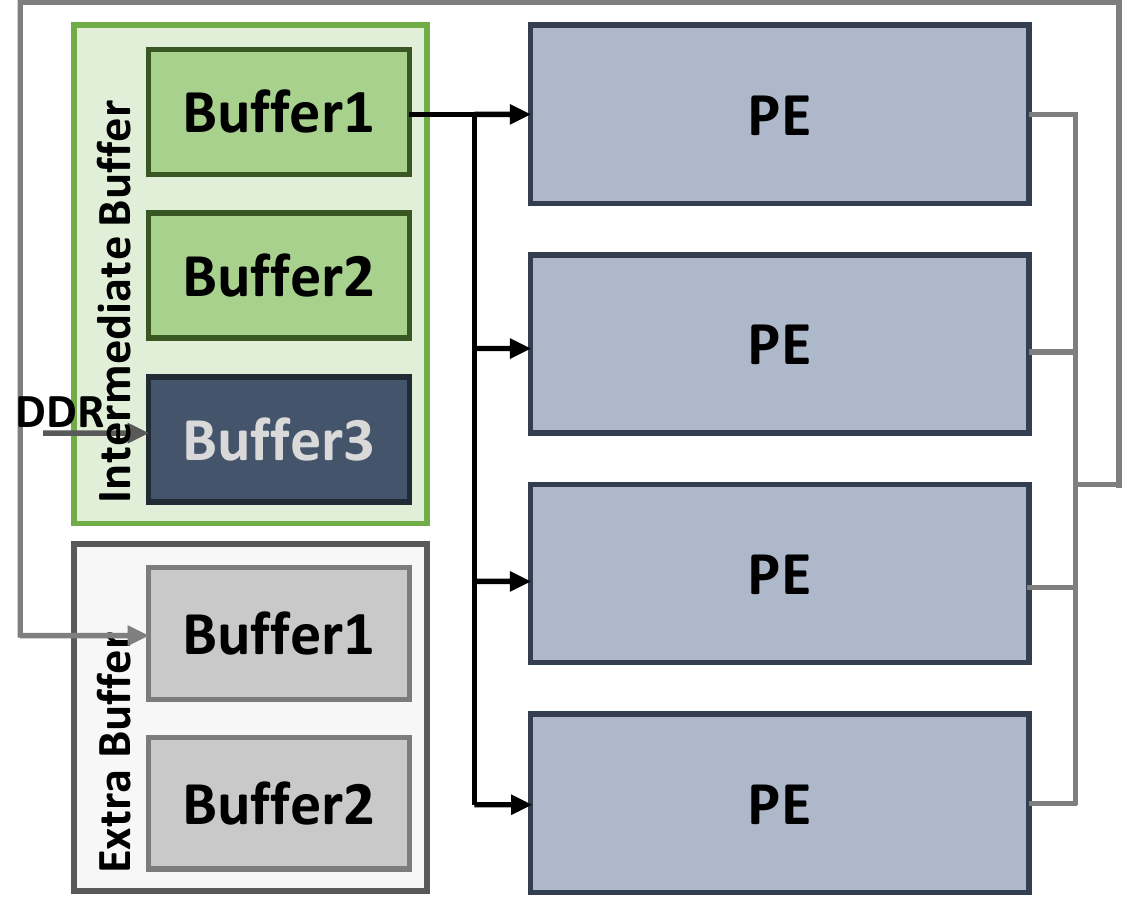}
\end{minipage}}
\subfigure[CONV1 for block 2]{
\begin{minipage}[b]{0.3\textwidth}
\centering
\includegraphics[width=\textwidth]{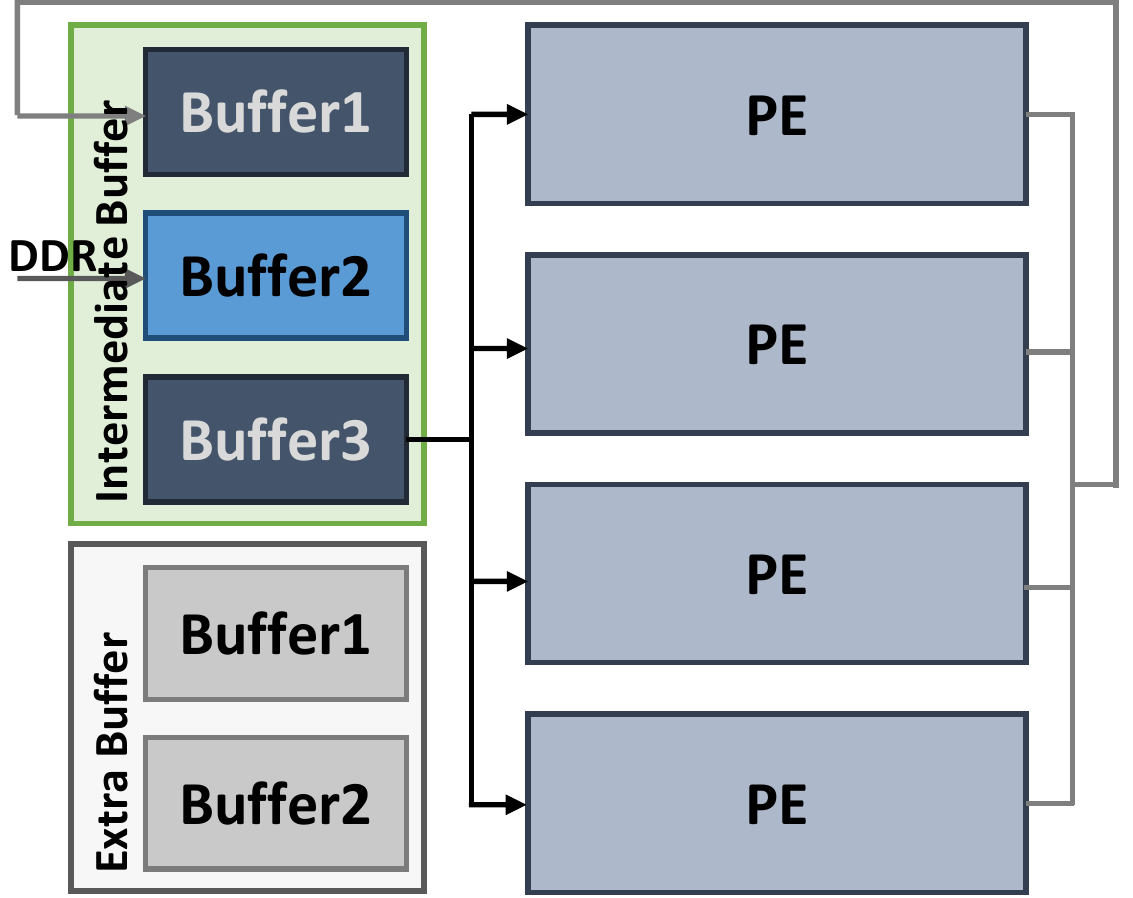}
\end{minipage}}
\subfigure[CONV4]{
\begin{minipage}[b]{0.292\textwidth}
\centering
\includegraphics[width=\textwidth]{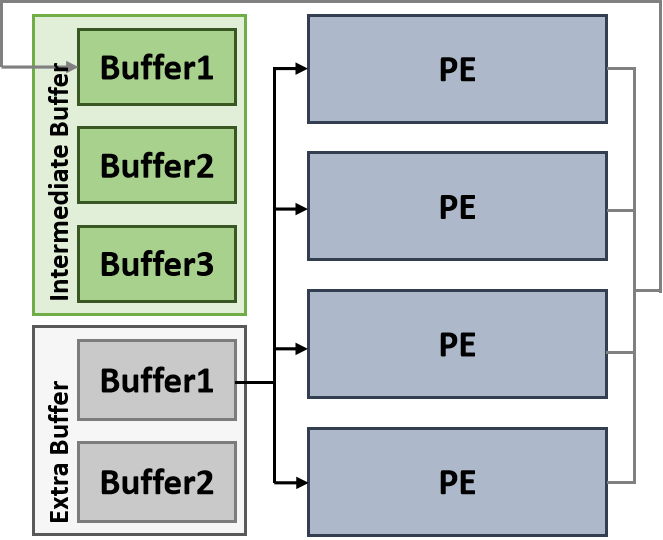}
\end{minipage}}
\caption{An illustration of memory organization and computing flow of three consecutive block convolutional layers and a normal convolutional layer. (b) In the beginning, block 1 (green) is read from Intermediate Buffer 1, and store the CONV1 results in Intermediate Buffer 2. At the same time, the accelerator starts to load block 2 (grey) from off-chip DRAM to Intermediate Buffer 3. (c) When CONV1 for block 1 is finished, the Intermediate Buffer 2 is served as input to calculate CONV2 for block 1, and store the results back to Intermediate Buffer 1 for subsequent computing of CONV3. (d) The CONV3 results of block 1 are stored in the Extra Buffer rather than another Intermediate Buffer, since fusion occurs in the output of CONV3, and larger buffers are needed. (e) When the CONV3 results of block 1 are obtained, the processing of block 2 is carried out in the same manner. (f) Results of CONV3 are concatenated in the Extra Buffer, and the output of CONV4 are stored in the Intermediate Buffer.}
\label{fig10}
\end{figure*}

\begin{figure}[]
\centering
\subfigure{
\includegraphics[width=0.465\columnwidth]{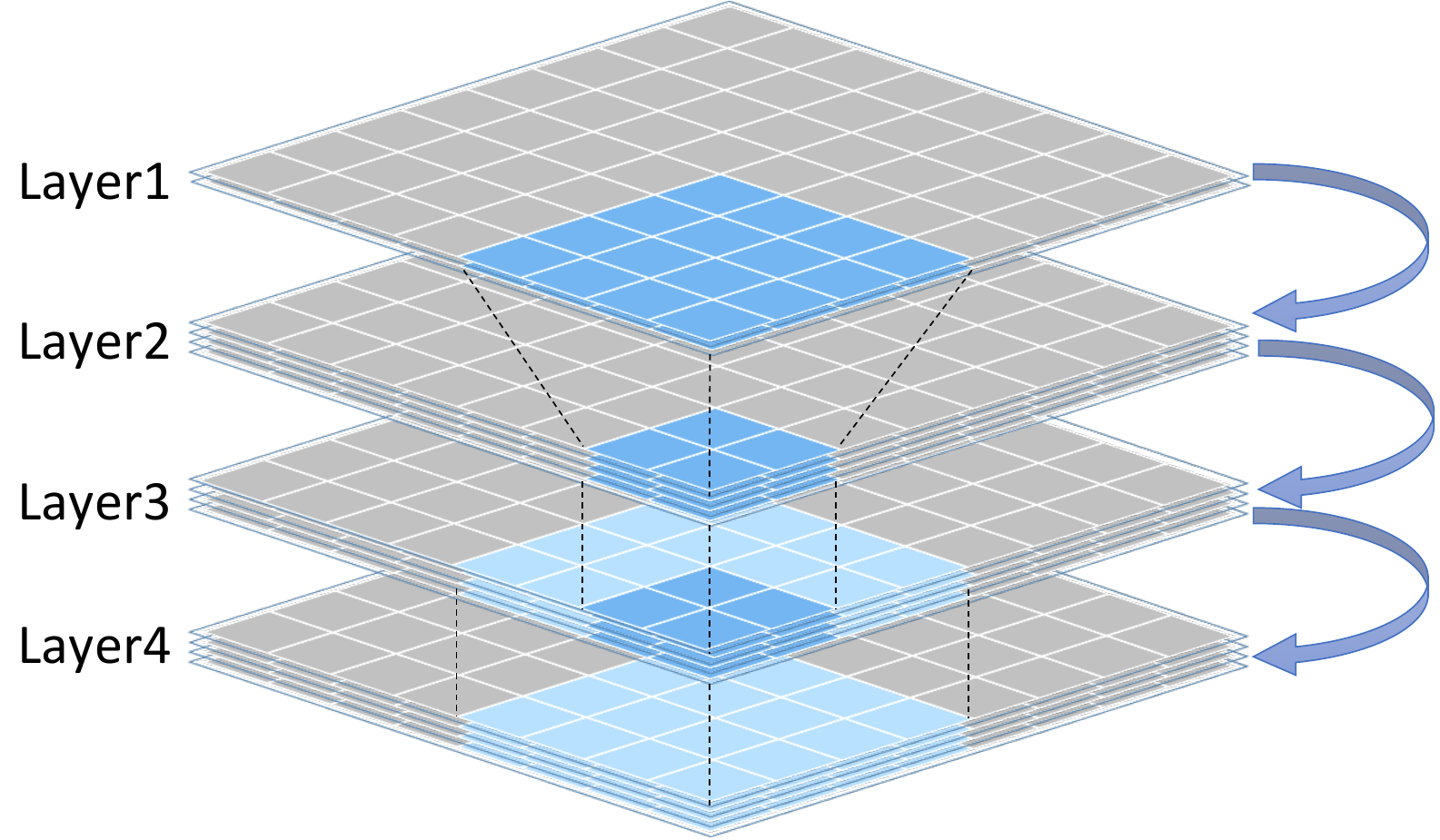}
}
\subfigure{
\includegraphics[width=0.465\columnwidth]{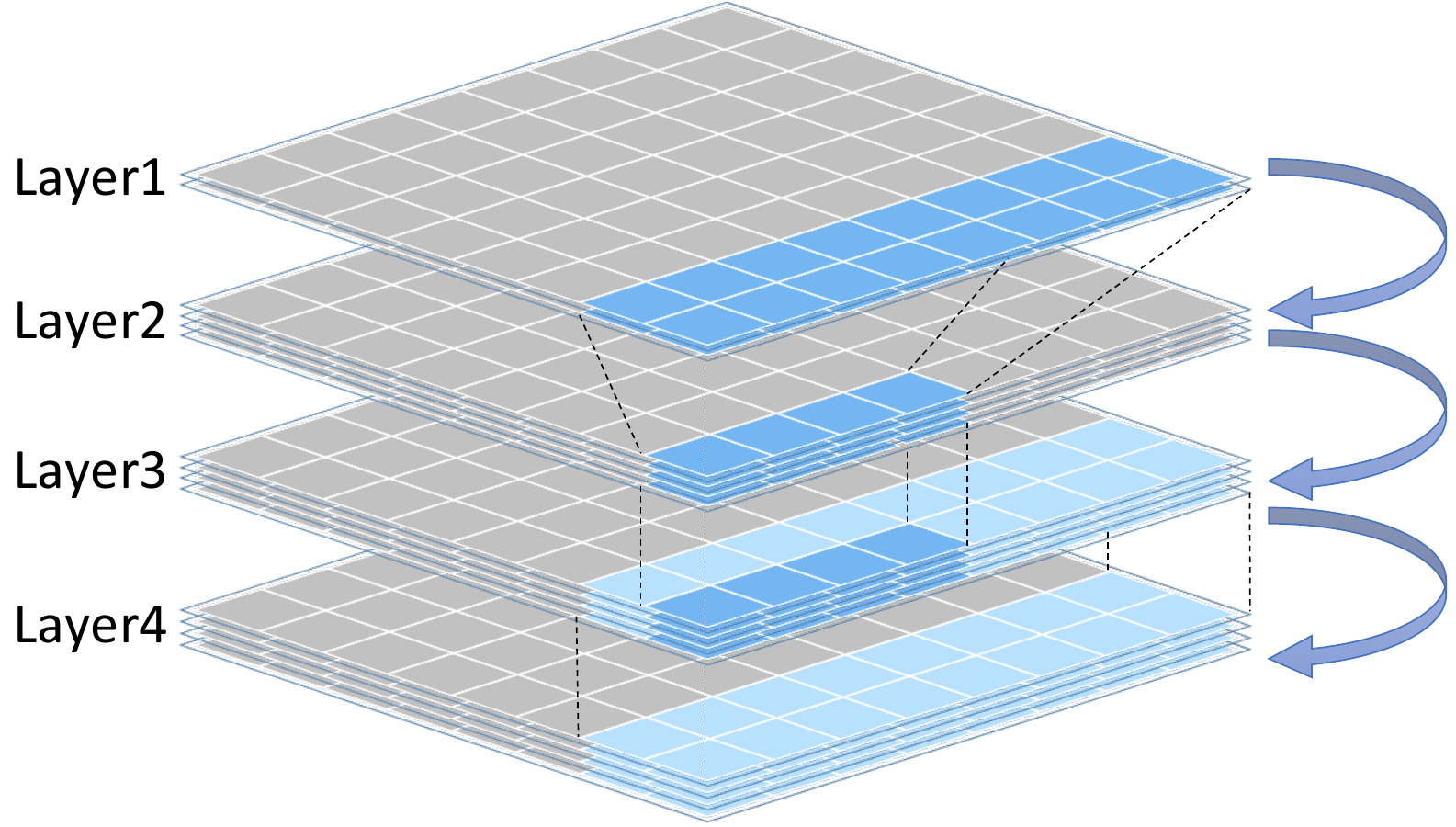}
}
\caption{Example of squared and non-squared layer fusion under fixed blocking.}
\label{fig11}
\end{figure}

\subsection{Application 1: VGG-16 Based Image Classification}
\subsubsection{Baseline}
\label{base}

We build our accelerator for VGG-16 based on the design in \cite{qiu2016going}, a state-of-the-art FPGA-based CNN accelerator. It employs loop optimization, such as loop unrolling, tiling, and interchanging to organize the computation for high throughput. 
Listing~\ref{listing1} shows the pseudo code of the dataflow. With loop tiling, the computation in each layer is conducted by invoking the PE multiple times. In each phase, the tiled input feature maps and the  corresponding filters are loaded from external DRAM to on-chip buffers. When the calculation is done, the partial output results are transferred back to DRAM. The inner-most three loops are fully unrolled for high parallelism.

\subsubsection{Multi-Layer Fusion Based on Block Convolution}
As illustrated in Section~\ref{algo}, block convolution enables efficient convolution by pruning away the massive data dependency at the tile boundaries, which is well structured for hardware acceleration.
Figure~\ref{fig11} shows how multiple blocked convolutional layers and pooling layers are fused during inference. 
It consists of four consecutive layers. Layer 1 is the input, which contains four input blocks of size $4\times 4\times C_{in}$. After a $2\times2$ pooling operation, the size of the output blocks in Layer 2 is only a quarter of the input's. With block convolution, the obtained blocks in Layer 2 can be immediately used for calculating the blocks in Layer 3 without additional data. In this scenario, we can allocate small buffers to store the output blocks in Layer 2, which is not possible in previous works. The operation from Layer 3 to Layer 4 is a little different. The size of the output blocks in Layer 4 is four times larger than that in Layer 3. Therefore, a larger data buffer is needed to caching multiple output blocks in Layer 3. Note that this will pose little pressure on the storage since the resolution of feature maps are getting lower as the network deepens.

Note that square blocking in the spatial dimension ($T_{r}=T_{c}$) may cause memory underutilization.
Suppose a single channel feature map of size $128\times128$, when square blocking is utilized, it can be partitioned into the size of $128\times128, 64\times64, 32\times32, 16\times16$, etc. If the on-chip memory capacity is $128\times100$, the largest block that can fit on chip, in this scenario, is the one of size $64\times64$, thus the memory utilization is only 40.96\%.
However, If we use rectangular blocking, such as $128\times64$, the memory utilization can be easily doubled. Consequently, we take rectangular blocking into consideration in our multi-layer fusion design to further improve the memory efficiency, as shown in Figure~\ref{fig11}.

\begin{table*}[]
	\centering
	\caption{Fused-Layer configurations of convolutional layers for VGG-16. The numbers in the second row indicate the grouping style. Each pair of numbers stand for the blocking sizes in width and height dimension [$T_{r}, T_{c}$].}
	\resizebox{\textwidth}{!}{
		\begin{tabular}[h]{|c|c|c|c|c|c|c|c|}
			\hline
			&\textbf{A}&\textbf{B}&\textbf{C}&\textbf{D}&\textbf{E}&\textbf{F}&\textbf{G} \\
			\cline{2-8}
			& \text{ \quad 2, 2, 3, 3, 3 \quad } & \text{ \quad 2, 5, 3, 3 \quad }  & \text{ \quad 2, 2, 3, 3, 3 \quad } &\text{ \quad 2, 2, 3, 3, 3 \quad }&\text{ \quad 2, 2, 3, 3, 3 \quad }&\text{ \quad 2, 2, 3, 3, 3 \quad }&\text{ \quad 2, 2, 3, 5 \quad}\\
			\hline
			\textbf{ \quad conv1-1 \quad } & [ 14, 14 ] & [ 28, 28 ] & [ 28, 28 ] & [ 14, 14 ] & [ 28, 28 ] & [ 28, 28 ] & [ 28, 28 ] \\
			\cline{1-1}
			\textbf{conv1-2} &[ 14, 14 ]&[ 28, 28 ]&[ 28, 28 ]&[ 14, 14 ]&[ 28, 28 ]&[ 28, 28 ]&[ 28, 28 ] \\
			\hline
			\textbf{conv2-1} &[ 14, 14 ]&[ 28, 28 ]&[ 28, 28 ]&[ 14, 14 ]&[ 28, 28 ]&[ 28, 28 ]&[ 28, 28 ] \\
			\cline{1-1}
			\textbf{conv2-2} &[ 14, 14 ]&[ 28, 28 ]&[ 28, 28 ]&[ 14, 14 ]&[ 28, 28 ]&[ 28, 28 ]&[ 28, 28 ] \\ \cline{1-2} \cline{4-8}			
			\textbf{conv3-1} &[ 14, 14 ]&[ 14, 14 ]&[ 28, 14 ]&[ 14, 14 ]&[ 28, 14 ]&[ 28, 28 ]&[ 28, 28 ] \\
			\cline{1-1}
			\textbf{conv3-2} &[ 14, 14 ]&[ 14, 14 ]&[ 28, 14 ]&[ 14, 14 ]&[ 28, 14 ]&[ 28, 28 ]&[ 28, 28 ] \\
			\cline{1-1}
			\textbf{conv3-3} &[ 14, 14 ]&[ 14, 14 ]&[ 28, 14 ]&[ 14, 14 ]&[ 28, 14 ]&[ 28, 28 ]&[ 28, 28 ] \\
			\hline
			\textbf{conv4-1} &[ 14, 14 ]&[ 14, 14 ]&[ 14, 14 ]&[ 14, 14 ]&[ 14, 14 ]&[ 28, 14 ]&[ 28, 28 ] \\
			\cline{1-1}
			\textbf{conv4-2} &[ 14, 14 ]&[ 14, 14 ]&[ 14, 14 ]&[ 14, 14 ]&[ 14, 14 ]&[ 28, 14 ]&[ 28, 28 ] \\
			\cline{1-1}
			\textbf{conv4-3} &[ 14, 14 ]&[ 14, 14 ]&[ 14, 14 ]&[ 14, 14 ]&[ 14, 14 ]&[ 28, 14 ]&[ 28, 28 ] \\
			\cline{1-7}
			\textbf{conv5-1} &[ 14, 14 ]&[ 14, 14 ]&[ 14, 14 ]&[ 14, 14 ]&[ 14, 14 ]&[ 14, 14 ]&[ 14, 14 ] \\
			\cline{1-1}
			\textbf{conv5-2} &[ 14, 14 ]&[ 14, 14 ]&[ 14, 14 ]&[ 14, 14 ]&[ 14, 14 ]&[ 14, 14 ]&[ 14, 14 ] \\
			\cline{1-1}
			\textbf{conv5-3} &[ 14, 14 ]&[ 14, 14 ]&[ 14, 14 ]&[ 14, 14 ]&[ 14, 14 ]&[ 14, 14 ]&[ 14, 14 ] \\
			\hline			
		\end{tabular}	
	}
	\label{tab6}
\end{table*}

\subsubsection{Dataflow and Memory System}
The baseline accelerator allocates two buffers for input and output data. In each invoking of the PE, a batch of tiles are loaded into the input buffer and store the intermediate results to the output buffer for off-chip transfer. Differently, we divide the data buffer into two \emph{intermediate buffers} and multiple \emph{extra buffers}, where the intermediate buffers are used for storing all the reusable intermediate results, and the extra buffers are used for caching additional data for larger blocks when fixed blocking is used. In each phase, the PEs read input blocks from one of the intermediate buffers, and save the partial results to another intermediate buffer that is served as the input later on. An example of dataflow and data buffer organization is shown in Figure~\ref{fig10}.

\subsubsection{Design Space Exploration}
\label{dse}
Since different blocking strategies may lead to significant variance in performance, in this section, we explore the design space according to inference latency and on-chip memory consumption to find the best trade-off.

\textbf{Inference Latency.}
Suppose the input and output tensors are of size $N\times R\times C$ and $M\times R\times C$, the corresponding blocking size is ($T_{n}, T_{r}, T_{c}$) and ($T_{m}, T_{r}, T_{c}$), and the number of PE is $N_{pe}$. Therefore, the computational cycles of a convolutional layer according to Section~\ref{base} can be determined as:
\begin{equation}
  Cycle_{conv}=\frac{N\times (T_{r}+2)\times (T_{c}+2)\times T_{m}}{N_{pe}}
\end{equation}
where, 
\begin{equation}
  N=\left \lceil \frac{M}{T_{m}} \right \rceil \times \left \lceil \frac{N}{T_{n}} \right \rceil \times \left \lceil \frac{R}{T_{r}} \right \rceil \times \left \lceil \frac{C}{T_{c}} \right \rceil
\end{equation}
is the total number of computational phases of a layer.

\begin{figure}
\centering
\subfigure[16-bit quantization, 2 PEs]{
\begin{minipage}[b]{0.9\columnwidth}
\centering
\includegraphics[width=\textwidth]{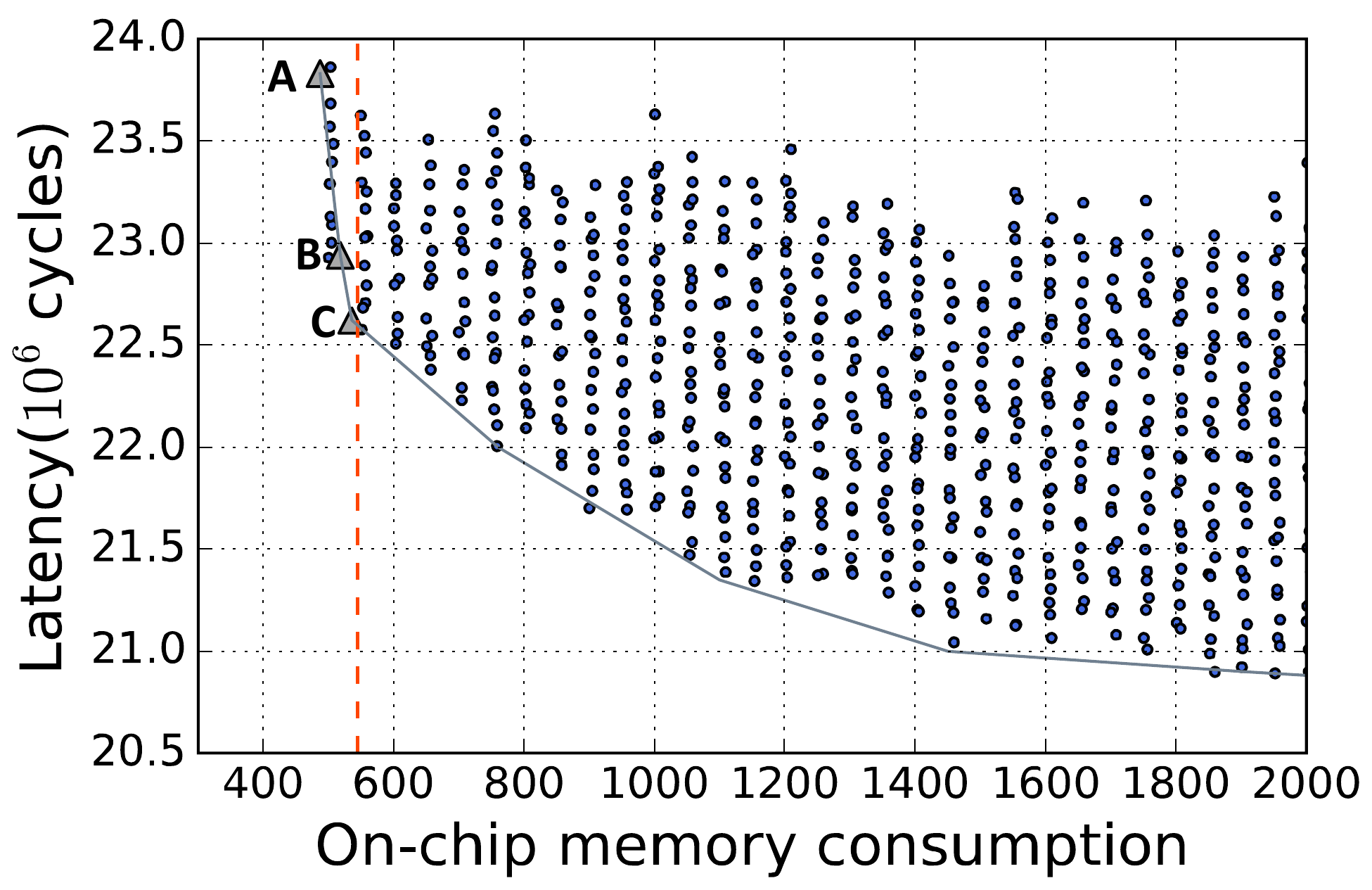}
\end{minipage}}
\subfigure[8-bit quantization, 4 PEs]{
\begin{minipage}[b]{0.9\columnwidth}
\centering
\includegraphics[width=\textwidth]{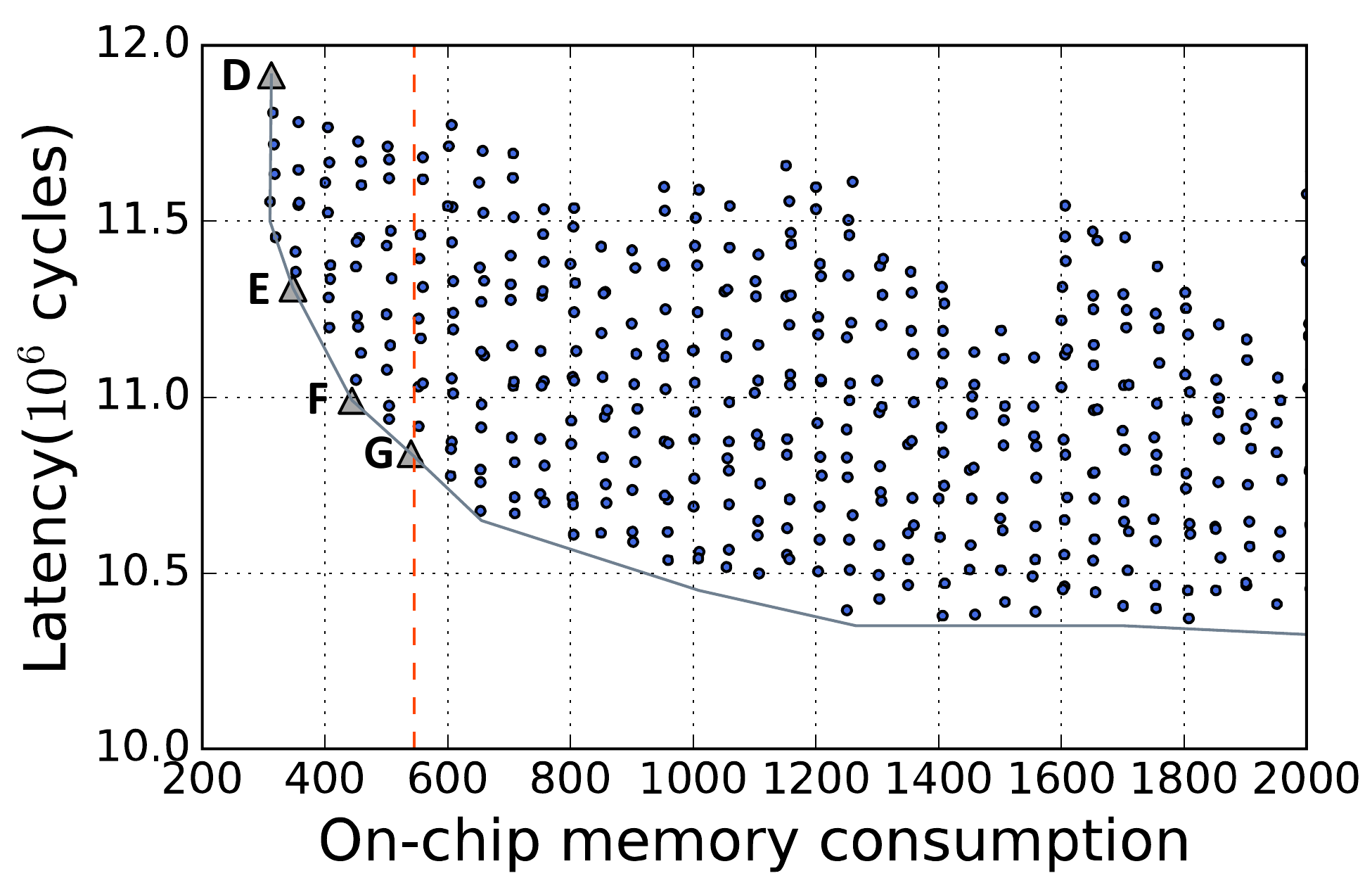}
\end{minipage}}
\caption{Design space exploration according to inference latency and on-chip memory consumption. The x-axis refers to the BRAM consumption of a specific fusing style. The dotted red line indicates the total amount of available on-chip BRAM of Xilinx Zynq ZC706 SoC. Points to the left of the dotted line stand for solutions that can accommodate all the intermediate data on-chip during inference.}
\label{fig12}
\end{figure}

\textbf{On-Chip Memory Consumption.}
For a given network structure, there are a number of ways to fuse layers into groups. For example, a three-layer network $\{l_{1}, l_{2}, l_{3}\}$ can be divided into multiple groups like $l_{1}-l_{2}-l_{3}$, $l_{1}-l_{2}l_{3}$, $l_{1}l_{2}-l_{3}$, and $l_{1}l_{2}l_{3}$. Each grouping style has a corresponding on-chip storage requirement.
Figure~\ref{fig12} shows all the fusion configurations for VGG-16. We explore the design space using a brute-force manner, each point in the figure represents a specific fusion strategy with corresponding theoretical inference latency and on-chip memory consumption. For each design point, we take blocking style, blocking size into consideration to derive the latency, and the on-chip BRAM consumption is based on the estimation from Vivado reports. We also consider the amount of PEs and quantization bitwidth to further extend the design space.

\begin{table*}[]
	\centering
	\caption{Comparison with state-of-the-art FPGA accelerators for VGG-16.}
		\begin{tabular}[h]{|c|c|c|c|c|c|c|c|c|}
			\hline
			                      &   \textbf{\multirow{2}{*}{\cite{qiu2016going}}} &     \textbf{\multirow{2}{*}{\cite{suda2016throughput}}}    &  \textbf{\multirow{2}{*}{\cite{zhang2018caffeine}}}   &  \textbf{\multirow{2}{*}{\cite{zhang2017frequency}}}  &   \textbf{\multirow{2}{*}{\cite{ma2017optimizing}}}  & \textbf{\multirow{2}{*}{\cite{fpgacluster}}}  &\textbf{\multirow{2}{*}{\cite{opu}}} &  \textbf{\multirow{2}{*}{Ours}} \\
			                       &     &         &     &        &      &  &     &      \\
			\hline
\textbf{\multirow{2}{*}{Platform}} &         Zynq         &                     Stratix-V   &        Virtex-7        &     Intel QuickAssist    &        Arria-10        &   Virtex-7   & Zynq &   Zynq \\
                                   &         ZC706        &                       GSD8         &         VX690t         &          QPI FPGA        &         GX1150         &   VX690t  & XC7Z100 &  ZC706   \\
			\hline
\textbf{\multirow{2}{*}{Precision}}                 &     \multirow{2}{*}{16bits fixed}     &   \multirow{2}{*}{8-16bits fixed}   &   \multirow{2}{*}{16bits fixed}   &   \multirow{2}{*}{32bits float}  &   \multirow{2}{*}{8-16bits fixed}  &\multirow{2}{*}{16bits fixed} &\multirow{2}{*}{8bits fixed} &\multirow{2}{*}{8bits fixed} \\
            &&&&&& &&\\
			\hline
            \textbf{\multirow{2}{*}{Technology}}    &  \multirow{2}{*}{28nm}          &   \multirow{2}{*}{28nm}          &         \multirow{2}{*}{28nm}            &        \multirow{2}{*}{28nm}              &            \multirow{2}{*}{20nm}       &\multirow{2}{*}{28nm} &\multirow{2}{*}{28nm}&\multirow{2}{*}{28nm}\\
            &&&&&&& &\\
            \hline
\textbf{Frequency}                 & \multirow{2}{*}{150}&     \multirow{2}{*}{120}  &  \multirow{2}{*}{150}  &    \multirow{2}{*}{200}  &  \multirow{2}{*}{150} & \multirow{2}{*}{150}& \multirow{2}{*}{200}&\multirow{2}{*}{150}\\	
            \textbf{(MHz)}&&&&&&&&  \\
			\hline
			\textbf{\multirow{2}{*}{BRAMs}}         &  \multirow{2}{*}{1090$\times$18k}    &      \multirow{2}{*}{2567$\times$20k}      &     \multirow{2}{*}{2940$\times$18k}          &   \multirow{2}{*}{2560$\times$20k}             &  \multirow{2}{*}{2713$\times$20k}   & \multirow{2}{*}{2940$\times$18k}& \multirow{2}{*}{1510$\times$18k}& \multirow{2}{*}{1090$\times$18k} \\
            &&&&&&&& \\
			\hline
			\textbf{\multirow{2}{*}{DSPs}}          &          \multirow{2}{*}{900}         &                \multirow{2}{*}{1963}         &    \multirow{2}{*}{3600}          &           \multirow{2}{*}{512}           &   \multirow{2}{*}{1518}          & \multirow{2}{*}{3600} &\multirow{2}{*}{2020} &\multirow{2}{*}{900} \\
            &&&&&&&& \\
			\hline
\textbf{Performance}   &   \multirow{2}{*}{136.97}&   \multirow{2}{*}{117.8}  &  \multirow{2}{*}{354} & \multirow{2}{*}{123.48}$^{\mathrm{a}}$ & \multirow{2}{*}{645.25}& \multirow{2}{*}{203.9} &\multirow{2}{*}{354} &\multirow{2}{*}{374.98}\\
            \textbf{(GOP/s)}&&&&&&&&  \\
			\hline
            \textbf{Latency/Image}&\multirow{2}{*}{224.6} &\multirow{2}{*}{262.9}&\multirow{2}{*}{87.29}&\multirow{2}{*}{263.27}$^{\mathrm{a}}$&\multirow{2}{*}{47.97}& \multirow{2}{*}{151.8}&\multirow{2}{*}{88.65} &\multirow{2}{*}{82.03}\\
            \textbf{(ms)}&&&&&&&&  \\
            \hline
			\textbf{Intermediate}   &\multirow{2}{*}{Yes}&  \multirow{2}{*}{Yes}   &     \multirow{2}{*}{Yes}&   \multirow{2}{*}{Yes}   &   \multirow{2}{*}{Yes}   & \multirow{2}{*}{Yes} & \multirow{2}{*}{Yes} &\multirow{2}{*}{\bf No} \\
             \textbf{layer transfer} &                    &                               &                         &        &                  &                           &     &            \\
			\hline
            \multicolumn{9}{r}{$^{\mathrm{a}}$The original paper only reports results of convolutional layers.}  \\
		\end{tabular}	
	\label{tab7}
\end{table*}

As shown in Figure~\ref{fig12}, for both 8-bit and 16-bit quantization, there are many configurations to implement VGG-16 with a high speed and keep all the intermediate data staying on chip. We list some feasible configurations with each layer's blocking size ($T_{r}, T_{c}$) in Table~\ref{tab6}. We can notice that the blocking size of A and D is always $[14,14]$ through the entire network, which allows them to allocate smaller data buffers to reduce memory consumption. On the contrary, C and G achieve higher performance by using either rectangular blocking or hierarchical blocking to fuse more layers into one group, which significantly increases the on-chip memory utilization.

\subsubsection{Evaluation}
Our implementation of VGG-16 is built on the Xilinx Zynq ZC706 SoC development board which consists of an ARM dual-core CPU and a Kintex-7 FPGA with 19.1 Mbits on-chip BRAM.

\begin{figure}[]
\centerline{\includegraphics[width=\columnwidth]{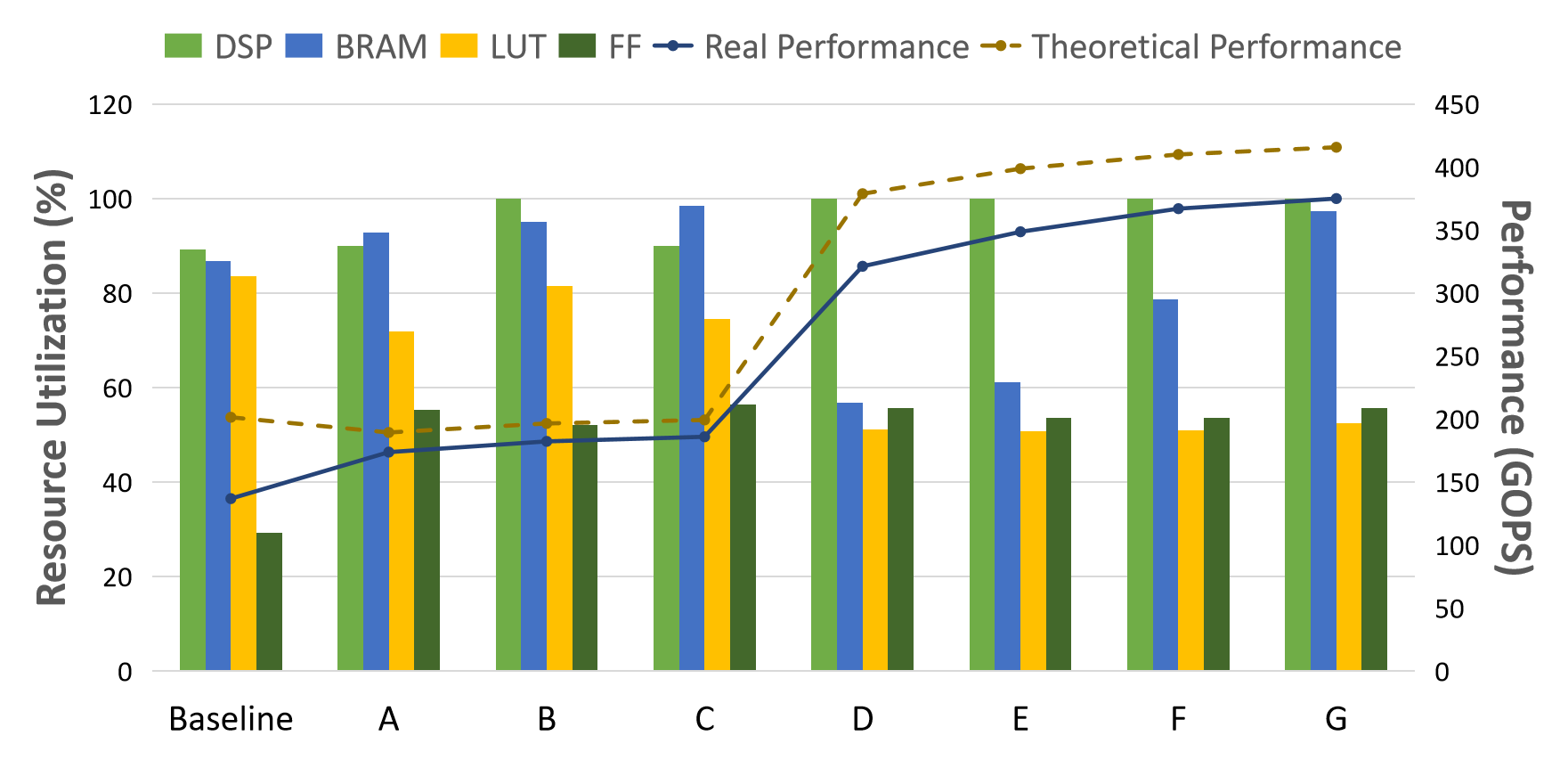}}
\caption{Comparison of our variant designs with the baseline in terms of on-chip resource utilization and performance. The
design of the baseline, A, B, C are based on 16-bit quantization with 2 PEs, and D, E, F, G are based on 8-bit quantization with 4 PEs.}
\label{fig13}
\end{figure}

We first evaluate the resource utilization and performance of our variant designs, as shown in Figure~\ref{fig13}. Compared with the baseline, although we move the intermediate data buffer completely from external DRAM to on-chip memory, there is only about 10\% increase on BRAM consumption. Moreover, it can be noticed that the real performance of our designs is also higher than the baseline. We attribute this to the following reasons: 1) With block convolution, all the network intermediate data can stay on chip during inference, which significantly reduces the DRAM access latency as well as CPU interrupts; 2) Rectangular blocking can further reduce the latency introduced by CPU interrupts compared to square blocking; 3) CPU interrupts caused by filter transfers can be significantly decreased by loading more filters at a time.
From Figure~\ref{fig13}, we can also notice that there is a gap between theoretical performance and real performance. This is mainly caused by frequent CPU interrupts in the process of filter transfers. If we can enlarge the on-chip filter buffers to reduce the amount of data transfer, this gap is likely to be narrowed. However, this happens at the cost of extra on-chip memory requirements.

Next, we compare our design G in Figure~\ref{fig13} with existing works, as shown in Table~\ref{tab7}. It can be seen that our solution achieves the highest performance among 28nm FPGAs with the minimum amount of on-chip memory. Although the better performance on more advanced FPGA can be obtained by utilizing more memory and DSP \cite{ma2017optimizing}, we are the first to deploy the very large VGG-16 network on low-cost FPGA without buffering intermediate data on external DRAM during inference.

\subsection{Application 2: VDSR Based Single Image Super-Resolution}
\subsubsection{Baseline}
The network structure of VDSR is shown in Table~\ref{tab8}. It is composed of 20 standard $3\times 3$ convolutional layers. Here we examine the single image super-resolution application with an input size of $1920\times 1080$. We design a DaDianNao-like \cite{luo2016dadiannao} accelerator as the baseline. All the weights and activations are quantized to 4 bits and 8 bits, respectively. 
Note that the volume of intermediate feature maps in each layer is 126.6 MB except for the last layer, which far exceeds the available on-chip memory of off-the-shelf FPGA. Therefore, tiling is used to partition the feature maps of a layer into several overlapped data blocks in the spatial dimension.

\begin{table}[]
\caption{VDSR architecture.}
\label{tab8}
\resizebox{\columnwidth}{!}{%
\begin{tabular}{c|c|c}
\hline
\multicolumn{1}{c|}{Type / Stride / Padding} & \multicolumn{1}{c|}{Filter Shape} & Input Size  \\ \hline
Conv / 1 / 1                             & $3\times 3\times 1\times 64$                          & $1080\times 1920\times 1$ \\ \hline
18 $\times$ Conv / 1 / 1                             & $3\times 3\times 64\times 64$                          & $1080\times 1920\times 64$\\ \hline
Conv / 1 / 1                             & $3\times 3\times 64\times 1$                           & $1080\times 1920\times 64$\\ \hline
Eltwise sum                                             & -                                 & $1080\times 1920\times 1$ \\\hline
\end{tabular}%
}
\end{table}

The computing core of the baseline accelator contains 8 PEs designed for vector-vector dot product between weights and activations along the channel dimension. Each PE contains 64 8/4-bit MAC, undertaking the calculation of one output channel. 
In addition to the computing core, the accelerator also contains data buffers for input and intermediate feature maps, a weight buffer that holds all the network weights, a DMA engine for high-throughput data transfer, and a global controller. We partition the feature maps into small tiles of size $27\times 48$ in the spatial dimension. Besides, all the data buffers are double buffered to hide the time for off-chip data transfer.
For each layer's computing, the accelerator iteratively loads a input tile to on-chip data buffer for calculation, and write an output tile to the main memory at the same time. Due to the dependency at the tile boundary, the output tiles should be either rearranged in the main memory or accessed in a scatter-gather manner before the next layer's processing. 

\subsubsection{End-to-end Layer Fusion Based on Block Convolution}
With block convolution, the data dependency among spatial tiles is eliminated. Therefore, for each tile of the input image, the calculation can be carried out layer by layer until the last layer. In this scenario, the transfer of intermediate feature maps is completely avoided during inference.
The workflow of our accelerator based on block convolution is as follows. Before the calculation starts, all the network weights are loaded into the on-chip weight buffer. When the start command is issued by the host CPU, a $27\times 48$ tensor is loaded into the input buffer for calculation. The output results are stored in the intermediate buffer and serve as the input to the next layer. The rest can be done in the same manner. When one of the output tile in the last layer is obtained, it is transferred to main memory, and the next input tile is loaded to on-chip buffer at the meantime. Note that in this scenario, double buffer is no longer needed since the bandwidth requirement is very low, off-chip data transfer only occurs in the first and the last layer. 

\subsubsection{Evaluation}
In this experiment, we implement both the baseline accelerator and the block convolution variant using HLS and synthesize the generated RTL in Vivado 2018.3. The resource consumption is shown in Table~\ref{tab9}. We choose the Xilinx Ultra96 MPSoC as our target platform, and both accelerators runs at 200MHz.
It can be seen that when block convolution is applied to the baseline accelerator, the consumption of BRAM is reduced. This is because the input and intermediate buffers no longer requiring ping-pong operation.
The amount of off-chip feature map transfer is drastically reduced by over 99.9\%, which not only reduces most of the energy consumption caused by accessing DRAM during inference, but also greatly alleviates the bandwidth pressure of DRAM. In addition, the data rearrangement in the main memory is completely avoided, which significantly improves the overall efficiency of the system.

\begin{table}[]
\caption{Resouce utilizaition and off-chip feature map transfer size of VDSR baseline accelerator and its block convolution variant.}
\label{tab9}
\resizebox{\columnwidth}{!}{
\begin{tabular}{c|c|c|c|l|c}
\hline
               & BRAM & LUT & FF & DSP & Transfer Size \\ \hline 
Baseline       &  352/432    & 69168/70560 & 4890/141120  &  265/360  &   36481.64 Mbits                       \\ \hline
Baseline+BConv &  264/432    &  69316/70560  &  4912/141120  &   265/360  &  31.64 Mbits                      \\ \hline
\end{tabular}
}
\end{table}

\section{Conclusion}
\label{conc}
In this paper, we propose block convolution, a hardware-friendly, simple yet efficient convolution operation that can completely avoid off-chip transfer of intermediate feature maps during inference, especially for memory-limited FPGA. The basic idea of block convolution is to eliminate the dependency of spatial blocks to improve computing locality and density. Extensive experiments demonstrate that comparable or higher accuracy can be achieved with block convolution for image classification, object detection, and single image super-resolution. We also showcase two CNN accelerators via algorithm/hardware co-design based on block convolution on resource-constraint FPGA, evaluation results show that both accelerators substantially outperform the baseline, without off-chip transfer of intermediate layers during inference.

\section{Acknowledgement}
This work was supported in part by National Natural Science Foundation of China (No.61972396), National Key Research and Development Program of China (No. 2020AAA0103402), and the Strategic Priority Research Program of Chinese Academy of Sciences (No. XDA27040300, XDB32050200).

\ifCLASSOPTIONcaptionsoff
  \newpage
\fi

\bibliographystyle{ieeetr}
\bibliography{ref}

\begin{IEEEbiography}[{\includegraphics[width=1in,height=1.25in,clip,keepaspectratio]{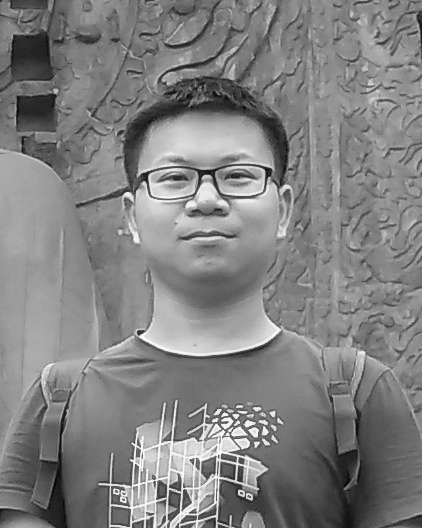}}]{Gang Li}
received the B.E. degree in information engineering from Northwestern Polytechnical University, Xi’an, China, in 2013. After that, he got the M.S. degree in National Laboratory of Pattern Recognition, Institute of Automation, Chinese Academy of Sciences in 2018. He is currently pursuing the Ph.D. degree from the Research Center for Brain-inspired Intelligence, Institute of Automation, Chinese Academy of Sciences. 

His research interests include deep learning,
computer architecture, and hardware/software co-design.
\end{IEEEbiography}

\vspace{-5 mm} 

\begin{IEEEbiography}[{\includegraphics[width=1in,height=1.25in,clip,keepaspectratio]{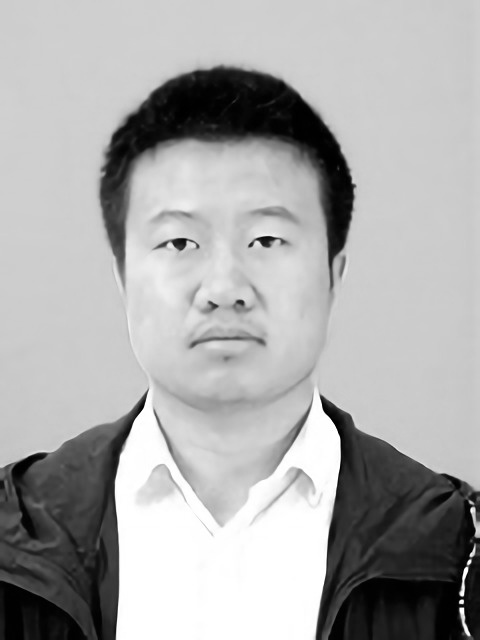}}]{Zejian Liu}
received the B.E degree in from Tianjin University, Tianjin, China, in 2018. He is currently working toward the PhD degree in the Institute of Automation, Chinese Academy of Sciences, Beijing, China. 

His current research interests include efficient algorithms and domain-specific hardware design. 
\end{IEEEbiography}


\begin{IEEEbiography}[{\includegraphics[width=1in,height=1.25in,clip,keepaspectratio]{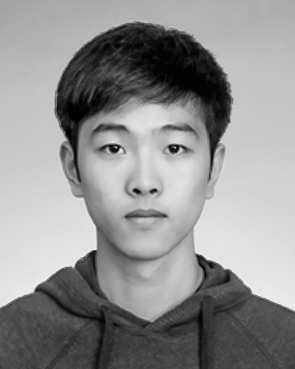}}]{Fanrong Li}
received the B.E. degree from Harbin Institute of Technology, Harbin, China, in 2017. He is currently pursuing the Ph.D. degree with the Institute of Automation, Chinese Academy of
Science, Beijing. 

His current research interests mainly focus on analyzing and designing computer architectures for
neural networks as well as developing heterogeneous reconfigurable accelerators.
\end{IEEEbiography}

\vspace{-165 mm} 

\begin{IEEEbiography}[{\includegraphics[width=1in,height=1.25in,clip,keepaspectratio]{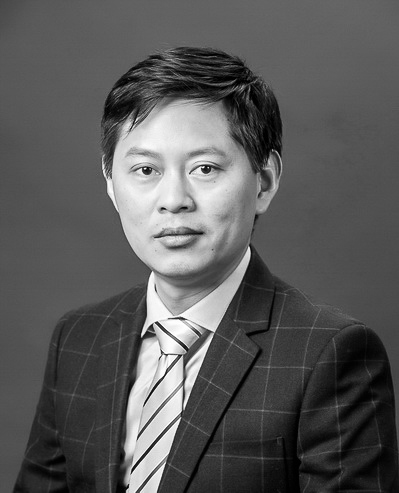}}]{Jian Cheng}
is a professor of Institute of Automation, Chinese Academy of Sciences. He received the B.S. and M.S. degrees in Mathematics from Wuhan University in 1998 and 2001, respectively. After that, he got Ph.D degree in pattern recognition and intelligent systems from Institute of
Automation, Chinese Academy of Sciences in 2004. 

His current major research interests include deep
learning, computer vision, chip design, etc.
\end{IEEEbiography}




\end{document}